\documentclass[twocolumn, aps, showpacs,preprintnumbers,bibnotes]{revtex4-2}
\pdfoutput=1
\usepackage[T1]{fontenc}

\usepackage{graphicx}
\usepackage{bm}
\usepackage[dvipsnames]{xcolor}
\usepackage[T1]{fontenc}
\usepackage{amsmath}
\usepackage{upgreek}
\usepackage{nicefrac}
\usepackage{multirow}
\usepackage{bbold}

\usepackage{hyperref}
\hypersetup{
pdfnewwindow=true, colorlinks=true,
linkcolor=blue, anchorcolor=blue,
citecolor=blue, filecolor=blue,
menucolor=blue, urlcolor=blue}

\makeatletter
\def\namedlabel#1#2{\begingroup
	#2%
	\def\@currentlabel{#2}%
	\phantomsection\label{#1}\endgroup
}
\makeatother

\begin{document}
	
	\title{Fragile topology in line-graph lattices with two, three, or four gapped flat bands}
	\author{Christie S. Chiu}
	\affiliation{
		Department of Electrical Engineering, Princeton University,
		Princeton, New Jersey, 08540, USA}
	\affiliation{
		Princeton Center for Complex Materials, Princeton University,
		Princeton, New Jersey, 08540, USA}
	\author{Da-Shuai Ma}
	\affiliation{
		Key Laboratory of Advanced Optoelectronic Quantum Architecture and Measurement (MOE), Beijing Key Laboratory of Nanophotonics, and Ultrafine Optoelectronic Systems, and School of Physics, Beijing Institute of Technology,
		Beijing 100081, China}
	\affiliation{
		Department of Physics, Princeton University,
		Princeton, New Jersey, 08540, USA}
	\author{Zhi-Da Song}
	\affiliation{
		Department of Physics, Princeton University,
		Princeton, New Jersey, 08540, USA}
	\author{B. Andrei Bernevig}
	\affiliation{
		Department of Physics, Princeton University,
		Princeton, New Jersey, 08540, USA}
	\author{Andrew A. Houck}
	\email{aahouck@princeton.edu}
	\affiliation{
		Department of Electrical Engineering, Princeton University,
		Princeton, New Jersey, 08540, USA}
	
	\date{\today}
	
		\begin{abstract}
		The geometric properties of a lattice can have profound consequences on its band spectrum.
		For example, symmetry constraints and geometric frustration can give rise to topologicially nontrivial and dispersionless bands, respectively.
		Line-graph lattices are a perfect example of both of these features: their lowest energy bands are perfectly flat, and here we develop a formalism to connect some of their geometric properties with the presence or absence of fragile topology in their flat bands.
		This theoretical work will enable experimental studies of fragile topology in several types of line-graph lattices, most naturally suited to superconducting circuits.
	\end{abstract}
	
	\maketitle
	
	\section{Introduction}
	
	Fragile topology is a property of a set of ``Wannier-obstructed'' gapped electronic bands whose Wannier obstruction can be resolved by adding select trivial bands \cite{Bradlyn2017, Po2018, Cano2018a, Bradlyn2019, BlancoDePaz2019, Hwang2019, Bouhon2019, Alexandradinata, Song2020a, JuanLuis2020, Else2019InteractingFragile, Liu2019shift, latimer2020correlated}.
	This Wannier obstruction refers to the inability to describe all states in these bands by exponentially localized symmetric Wannier functions, known as the atomic limit.
	``Extended'' states are then required, much like the edge states of topological insulators \cite{Kane2005TI,bernevig2006quantum,konig2007quantum,Fu2007TI,zhang2009topological,chen2009experimental,xia2009observation,kitaev2009periodic,schnyder2008classification}; crucially, however, the stable topology of these materials differs from fragile topology because it is robust to the addition of trivial bands.
	Additionally, the extended states of fragile phases generally do not exist at the edge.
	Recent theoretical and experimental work has found that fragile phases violate the bulk-boundary correspondence, but instead exhibit gapless edges under ``twisted'' boundary conditions \cite{Song2020, Peri2020}.
	Moreover, the fragile topology of electronic states also manifests itself in the contribution to the superfluid weight in the superconducting phase \cite{Xie2020topology,Hu2019TBG,Julku2020TBG,Hazra2019bounds} and the level crossings in Hofstadter spectrum under magnetic field \cite{Jonah2020Hof,lian2020landau}. 
	
	Fragile topology can also be characterized under the theory of topological quantum chemistry, which classifies topological bands by classifying all possible atomic limits based on crystallographic symmetries \cite{Po2017, Bradlyn2017, Kruthoff2017}.
	Under this theory, atomic limits are described by elementary band representations (EBRs) \cite{zak1980symmetry,zak1982band,Elcoro2017,Cano2018,Vergniory2017TQC}; while atomic bands can be written purely as a sum of EBRs, fragile topological bands cannot \cite{Po2018,Cano2018a}.
	Instead, they can be written as sums and differences of EBRs, such that the inclusion of trivial bands can render the entire set of bands trivial.
	In this work, we mainly focus on the so-called eigenvalue fragile states whose irreducible representations (irreps) in momentum space cannot be written as sums of EBRs.
	
	Less recently, theoretical work has also predicted that nearly flat bands with stable topology may give rise to fractional quantum Hall states at high temperatures or zero magnetic field \cite{Neupert2011, Wang2012, Regnault2011, Sun2011}.
	However, to our knowledge no exact flat bands with stable topology have been found in lattice models.
	On the other hand, fragile topological bands can be exactly flat, for example in magic-angle twisted bilayer graphene \cite{Bistritzer2011, Ahn2019, Po2019, Kang2019, Song2019, dai2019TBGLL}.
	For exact flatness, then, fragile topological bands provide an ideal platform for studies of strongly interacting quantum phases \cite{Leykam2018}.
	Recent works \cite{Kang2019, Bultinck2020, Lian2020, Zhang2019, Liu2019} have shown that the partially filled fragile-topological flat bands in twisted bilayer graphene could form various correlated insulating phases, including the Chern insulator phase, under different parameters.
	It has also been shown that, remarkably, the Chern insulator phase originates from the fragile topology, which allows a natural choice of the Chern band basis \cite{Bultinck2020, Hejazi2020, Song2020b,  Bernevig2020}. 
	
	Entire classes of lattices are known to have exactly flat bands, for example bipartite lattices with an unequal number of vertices in each part \cite{Lieb1989} or certain types of ``line-graph lattices'' \cite{Cvetkovic2004}.
	However, apart from directly computing the representation of specific flat-band systems, it is not generally known whether these bands are topological and, if so, whether the topology is stable or fragile.
	
	Here we consider line-graph lattices of ``regular'' lattices, defined by the attribute that every vertex has the same coordination number.
	The band spectra of these lattices have flat bands as their lowest energy bands.
	Although the topology of these bands can be computed via topological quantum chemistry, this must be done on a case-by-case basis.
	We develop a framework for analyzing the topology of line-graph-lattice flat bands for entire families of lattices, drawing connections between simple geometric attributes of the lattices and their flat-band representations.
	With this framework, we identify such families whose flat bands have fragile topology, as well as families of line-graph lattices whose flat bands are topologically trivial but that, after certain perturbations, can be split into fragile topological flat bands and topologically trivial dispersive bands.
	These results can inform experimental simulations of line-graph lattices for studies of fragile topology; in particular, these lattices are quite natural to simulate with coplanar waveguide resonators in quantum circuits because the line-shaped resonators act as lattice vertices for microwave photons, with tunneling between vertices made possible through capacitive coupling at the resonator ends \cite{Carusotto2020}.
	
	A line graph $L(X)$ can be formed from any graph $X$ (which we will refer to as the root graph) by placing a vertex $v_{L(X),i}$ on each edge $e_{X,i}$ of $X$ and connecting vertices $v_{L(X),i}$ and $v_{L(X),j}$ if their corresponding edges $e_{X,i}$ and $e_{X,j}$ are adjacent, i.e. share a common vertex.
	We then define the tight-binding Hamiltonian
	\begin{equation}
	\hat{H} = \sum_{\langle i,j \rangle} \hat{a}_i^\dagger \hat{a}_j + \hat{a}_j^\dagger \hat{a}_i
	\end{equation}
	where the sum is taken over all adjacent vertices $v_{L(x),i}$ and $v_{L(X),j}$, representing amplitude-1 hopping of spinless bosons $\hat{a}_i$ between adjacent vertices in the line graph.
	
	There are several properties of line graphs, discussed further in Appendix A of the Supplementary Material with examples \cite{SM}, that are relevant to this work:
	\begin{enumerate}
		\item[\namedlabel{itm:LGperiodic}{LG1}] If $X$ is a periodic lattice, $L(X)$ is as well.
		\item[\namedlabel{itm:LGspacegrp}{LG2}] Any symmetries of $X$ are inherited by $L(X)$, i.e. the space group of $X$ is the same as that of $L(X)$.
		\item[\namedlabel{itm:LGsubgraph}{LG3}] As a consequence of the line-graph construction, every vertex $v_{X, i}$ of the root graph gives rise to a ``complete subgraph'' in the line graph, where a complete subgraph is defined as a subset of $k$ vertices and binomial coefficient $\binom{k}{2}$ edges for which all pairs of vertices are connected by one of the edges (i.e. ``fully connected''). In these complete subgraphs, $k$ will be equal to the coordination number of $v_{X,i}$.
		\item[\namedlabel{itm:LGcycle}{LG4}] Consider a sequence of vertices of the root graph $(v_{X, 1}, v_{X, 2}, \dots v_{X, n+1})$ where $v_{X, 1}=v_{X, n+1}$ but all other vertices are distinct. Take the sequence of edges $(e_{X, 1}, e_{X, 2}, \dots e_{X, n})$ of $X$ where the edge $e_{X, i}$ connects vertices $v_{X, i}$ and $v_{X, i+1}$. These vertices and edges form a ``cycle'' of the graph. As a consequence of the line-graph construction, every cycle of $X$ gives rise to a cycle of equal length (number of edges) of $L(X)$. These cycles of $L(X)$ are ``chordless'', meaning that no two vertices of the cycle are connected by an edge that does not belong in the cycle.\label{LXcycle}
	\end{enumerate}
	For regular root-graph lattices $X$ with $n$ vertices per unit cell, each with coordination number (degree) $d$, we have additionally the following:
	\begin{enumerate}
		\item[\namedlabel{itm:LGspectrum}{LG5}] Given energies $E_X$ of $X$, its corresponding line-graph lattice $L(X)$ has energies $E_{L(X)} = \{E_X + d - 2, -2\}$, with one or more flat bands at $-2$.
		\item[\namedlabel{itm:LGdegen}{LG6}] The degeneracy $D$ of the flat band at $-2$ is given by $D = n(d-2)/2$.
		\item[\namedlabel{itm:LGgap}{LG7}] If $X$ is non-bipartite, then the flat band(s) at $-2$ for $L(X)$ will be gapped from the other bands.
	\end{enumerate}
	Finally, if $X$ (under periodic boundary conditions) can be embedded on a torus such that none of its edges cross each other, then we define the faces of $X$ to be regions bounded by edges and containing no edges or vertices.
	Because $X$ is on a torus, the coordination number $d$ and number of vertices $n$ per unit cell then determine the number of faces is per unit cell to be equal to the band degeneracy $D$:
	\begin{enumerate}
		\item[\namedlabel{itm:LGnumfaces}{LG8}] The number of faces per unit cell of $X$ is also given by $n(d-2)/2$.
	\end{enumerate}
	We consider line-graph lattices of non-bipartite toroidal regular root-graph lattices, with flat-band degeneracy $1 < D \leq 4$.
	These lattices have $C_2$, $C_3$, or $C_6$ symmetry, and can be further split into families based on their coordination number and the number of faces per unit cell that are bounded by an even number of edges (``even-sided faces'').
	We find that lattices in the same family have the same representation of the associated flat bands.
	More specifically, these three characteristics define which graph-element type---vertex, edge, or face---is located at each maximal Wyckoff position of the root-graph lattice unit cell.
	Maximal Wyckoff positions in a space group are the high-symmetry points in real space with the little groups---under which they are invariant---as maximal subgroups of the space group.
	Each element type (vertex, edge, or face) then determines the so-called real-space invariants (RSIs) of the flat-band at each maximal Wyckoff position, from which the representation and topology follow \cite{Song2020}.
	Furthermore, for $D=3$ and $D=4$ flat bands we consider various perturbations to reduce the degeneracy and identify a class of perturbations that produces fragile topological flat bands.
	
	\begin{figure}
		\includegraphics{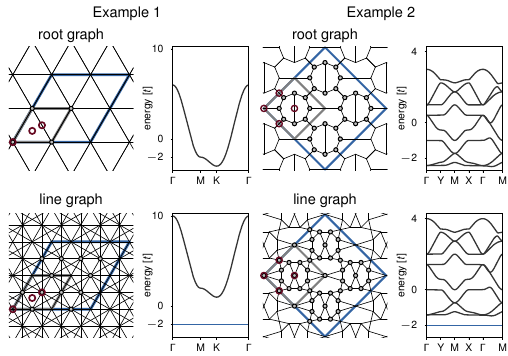}
		\caption{The two examples of line-graph lattices described in the main text. Example 1 begins with the triangle lattice as its root graph, and example 2 begins with the heptagon-heptagon-pentagon-pentagon lattice with $M_x$ and $M_y$ mirror symmetries. Upon taking the line graph of these root-graph lattices, the band spectra shift upward in energy (by $d-2$) and flat bands are created at $-2$. Unit cells are outlined in gray, lattice vertices in a two-unit-cell by two-unit-cell region (outlined in blue) are drawn as gray circles, and the maximal Wyckoff positions of one unit cell are drawn as red circles.}\label{fig:examples}
	\end{figure}
	
	In discussing our framework, we will use two elucidating examples, shown in Fig. \ref{fig:examples}; additional examples are included in Appendix F of the Supplementary Material \cite{SM}.
	For example 1 we take the line graph of the triangle lattice, which has coordination number 6, zero even-sided faces, and $C_6$ symmetry.
	It also has one vertex and two faces per unit cell; therefore, the corresponding line-graph lattice has a $D=2$-fold degeneracy of its flat bands at energy $-2$.
	For example 2 we take the line graph of the heptagon-heptagon-pentagon-pentagon lattice with $M_x$ and $M_y$ mirror symmetries as shown in Fig. \ref{fig:examples}.
	This root-graph lattice has coordination number 3, zero even-sided faces, and $C_2$ symmetry.
	It also has eight vertices and four faces per unit cell; therefore, the corresponding line-graph lattice has a $D=4$-fold degeneracy of its flat bands at $-2$.
	
	\section{From root-graph lattice properties to graph element at each maximal Wyckoff position}
	
	Maximal Wyckoff positions are labeled by a number according to their multiplicity and a letter defining their position (see top row of Table \ref{table:maxWyckpos}). They play a large role in the construction of EBRs.
	Previous works have considered which maximal Wyckoff positions are occupied by lattice vertices (atomic orbitals) to define EBRs \cite{Bradlyn2017,Elcoro2017,Cano2018,Vergniory2017TQC,Po2017,Kruthoff2017}. However, here we consider all graph elements of the lattice and whether maximal Wyckoff positions are occupied by vertices, edges, or faces of the root-graph lattice.
	In general, the lattices we consider contain many vertices on nonmaximal Wyckoff positions as well.
	As the first step in determining the properties under symmetry of the line-graph lattice flat band, we show the relationship between the root-graph lattice properties and the graph element at each maximal Wyckoff position.
	
	The maximal Wyckoff positions for our two examples are highlighted in Fig. \ref{fig:examples} as red circles.
	Example 1 has $C_6$ symmetry and its maximal Wyckoff positions are the $1a$, $2b$, and $3c$ positions, defined in Table \ref{table:maxWyckpos}.
	In its root-graph lattice (the triangle lattice), at the $1a$ position sits a vertex, at $2b$ is a face, and at $3c$ is an edge.
	As for example 2, its maximal Wyckoff positions are the $1a$, $1b$, $1c$, and $1d$ positions (Table \ref{table:maxWyckpos}) resulting from its $C_2$ symmetry.
	In its root-graph lattice, at all four are edges.
	
	More generally, we find a relationship between how many of each graph-element type are at a root-graph lattice's maximal Wyckoff positions, and the lattice's coordination number, number of even-sided faces, and symmetry.
	These correspondences are listed in Table \ref{table:maxWyckpos}, with cells pertaining to examples 1 and 2 colored in blue.
	Several patterns emerge across these root-graph lattices, stated and proved in Appendix C of the Supplementary Material \cite{SM}.
	
	\begin{table}[tb]
		\begin{centering}
			\begin{tabular}{ll@{\hspace{12pt}}c@{\hspace{10pt}}c@{\hspace{10pt}}ccc}
				\hline
				&&$C_2$& $C_3$& \multicolumn{3}{c}{$C_6$}\\
				&&\includegraphics{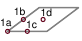}& \includegraphics{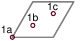}& \multicolumn{3}{c}{\includegraphics{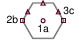}}\\
				&&$1a, 1b, 1c, 1d$&$1a, 1b, 1c$&$1a$&$2b$&$3c$\\
				\hline
				$D=2$&&-&2f, 1v&\color{RoyalBlue}{1v}&\color{RoyalBlue}{1f}&\color{RoyalBlue}{1e}\\
				$D=3$&\hspace{10pt}$d$ odd& 1f, 3e &\multirow{2}{*}{-}&1f&1f&1e\\
				&\hspace{10pt}$d$ even& 1f, 2e, 1v&&1f&1f&1v\\
				$D=4$&0 even faces&&&&\multirow{4}{*}{-}&\\
				&\hspace{10pt}$d$ odd&\color{RoyalBlue}{4e}&1f, 2v&&&\\
				&\hspace{10pt}$d$ even&4e OR 2e, 2v&&&&\\
				&2 even faces&2f, 2e&&&&\\
				\hline
			\end{tabular}
			\caption{For the maximal Wyckoff positions associated with a given point-group symmetry, depicted in the header row, we predict how many of them have vertices (v), edges (e), or faces (f) of the root-graph lattice at these positions based on the lattice's flat-band degeneracy $D$, coordination number $d$, and number of even-sided faces per unit cell.
				For example, the root-graph lattice of example 2 has $C_2$ symmetry, $D=4$, zero even faces, and $d$ odd, so the table indicates that its four maximal Wyckoff positions should be occupied by edges; indeed, as seen in Fig. \ref{fig:examples}, this is the case.
				We note that for the $C_3$- and $C_6$-symmetric lattices, we find a single lattice geometry for each cell in the table, drawn in Fig. S8 of \cite{SM}.
				Cells corresponding to examples 1 and 2 are in blue.}\label{table:maxWyckpos}
		\end{centering}
	\end{table}
	
	From the line-graph construction and properties \ref{itm:LGspacegrp}, \ref{itm:LGsubgraph}, and \ref{itm:LGcycle} of line graphs, we can determine which graph element of the line-graph lattice is on each of maximal Wyckoff positions, given which root-graph graph element is on each maximal Wyckoff position in the root graph.
	For example, as seen in Fig. \ref{fig:examples}, the triangle lattice's $1a$ maximal Wyckoff position is occupied by a vertex; $2b$ is occupied by a triangular face, which is bounded by a cycle of length 3; and $3c$ is occupied by an edge.
	Upon taking the line graph (see Appendix A of the Supplementary Material for details \cite{SM}), the root-graph vertex at $1a$ gives rise to a complete subgraph at $1a$ in the line graph, of six vertices that are pairwise fully connected by $\binom{6}{2} = 15$ edges (property \ref{itm:LGsubgraph} of line graphs).
	Similarly, the root-graph triangular face at $2b$ gives rise to a triangular face at $2b$ in the line graph (property \ref{itm:LGcycle}), and the root-graph edge at $3c$ gives rise to a vertex at $3c$ in the line graph (by definition of the line-graph construction).
	
	\section{From maximal Wyckoff position location type to real-space invariant}
	
	Real-space invariants (RSIs) are quantum numbers assigned to maximal Wyckoff positions and can be used to determine band topology.
	RSIs compute the local representation of an orbital at a Wyckoff position, which induces a set of bands in the Brillouin zone \cite{Song2020}.
	For a maximal Wyckoff position with point symmetry $C_s$, these eigenstates can have (single group) eigenvalues $e^{i2\pi k/s}$ for integer $k \in [0, 1, \dots s-1]$.
	Here we consider RSIs for two-dimensional point-group symmetries without spin-orbit coupling and with time-reversal symmetry (TRS).
	Due to TRS, there is a one-to-one correspondence between eigenstates with eigenvalue $e^{\pm i2\pi k/s}$, and hence we only consider $k \leq \lfloor s/2 \rfloor$.
	The RSIs at maximal Wyckoff position $w$ are then equal to the difference in multiplicities $m^s_{w, k\neq 0}$ and $m^s_{w, k=0}$ of these eigenstates: $\delta^s_{w, k'} \equiv m^s_{w, k'} - m^s_{w, 0}$ for $k' \in [1, s/2]$.
	We note that these RSIs are can also be written using the point group irreducible representation (orbital) notation from the Bilbao Crystallographic Server \cite{Elcoro2017}, but avoid this notation here for simplicity.
	
	A real-space approach to determine the RSIs of a $C_s$ center is to consider local energy eigenfunctions $|\phi\rangle$ plus each of their $C_s$ images with a relative phase:
	\begin{equation}
	|\phi_k\rangle \equiv |\phi\rangle + e^{i2\pi k/s}C_s|\phi\rangle + \dots + (e^{i2\pi k/s}C_s)^{s-1}|\phi\rangle \label{eq:CsEigenfn}
	\end{equation}
	Notice that each value of $k \in [0, s/2]$ generates an eigenfunction of eigenvalue $e^{i2\pi k/s}$.
	However, some of these constructions may yield $|\phi_k\rangle \propto |\phi\rangle$ (with an overall phase), which occurs when $|\phi\rangle$ is a $C_s$ eigenstate, or vanish identically.
	If either of these is the case, then one or more of the RSIs will nonzero-valued.
	To evaluate the RSIs for our line graphs, we choose a real-space flat-band eigenbasis containing so-called ``cycle'' and ``chain'' compact localized states (CLSes), which are defined in Appendix B of the Supplementary Material \cite{SM}.
	
	\begin{figure}[tb]
		\includegraphics[width=\columnwidth]{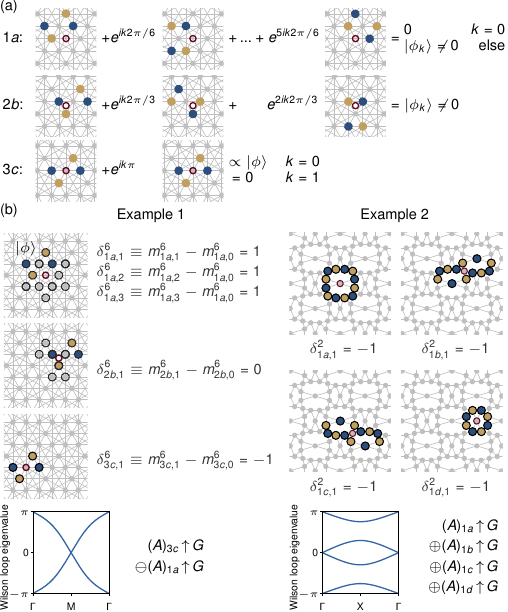}
		\caption{The point-group-symmetric eigenstates local to each maximal Wyckoff position (red circles) depict the real-space invariants (RSIs) for each position, as demonstrated by our two examples. Here, the local flat-band eigenfunctions $|\phi\rangle$ are based on compact localized states (CLSes, see main text). They are real-valued and depicted by the colored circles, with blue (yellow) circles denoting relative amplitude $+1$ ($-1$). 
			\textbf{(a)} $C_s$ flat-band eigenstate construction from flat-band energy eigenstates, as described in the main text, for each maximal Wyckoff position in example 1. The flat-band energy eigenstate $|\phi\rangle$ and its $C_s$ images are represented graphically.
			\textbf{(b)} RSI determination for examples 1 and 2 based on the multiplicities of $C_s$ flat-band eigenstates of each eigenvalue. Circles outlined in black highlight vertices where at least one of the $C_s$ images of $|\phi\rangle$ have nonzero amplitude. The representation follows directly from these RSIs, and we find odd Wilson loop winding when the representation involves a difference of EBRs as in example 1.}\label{fig:RSI}
	\end{figure}
	
	Figure \ref{fig:RSI} depicts the RSIs and associated CLS eigenstates at each maximal Wyckoff position for our two examples.
	For example 1, we define an flat-band eigenstate $|\phi\rangle$ with nonzero amplitude on four vertices in the line-graph lattice, enclosing an even cycle around two of the triangle faces.
	At the $1a$ position, we consider the sum of $|\phi\rangle$ with each of its $C_6$ images with a relative phase $e^{i2\pi k/6}$ [see Fig. \ref{fig:RSI}(a)].
	Of the integers $k \in [0, 3]$, all yield nonzero functions except for $k=0$.
	In particular, notice that the $C_s$ eigenstate constructions can vanish identically for some $k$ only if each vertex (of the line-graph lattice) where $|\phi\rangle$ has nonzero amplitude, also has nonzero amplitude for at least one of the $C_s$ images of $|\phi\rangle$.
	All other local flat-band symmetry eigenstates for the line graph of the triangle lattice involve a local energy eigenfunction $|\phi'\rangle$ that does not have this property; therefore, the constructions $|\phi'_k\rangle$ will construct the same number of eigenfunctions of each eigenvalue.
	Then the eigenstates $|\phi'_k\rangle$ do not contribute to the RSIs of the origin $1a$, and the RSIs are $\delta^6_{1a, 1} = \delta^6_{1a, 2} = \delta^6_{1a, 3} = 1$.
	The same procedure for the $2b$ and $3c$ positions yield RSIs of $\delta^6_{2b, 1} = 0$ and $\delta^6_{3c, 1} = -1$.
	
	In example 2, we define different local eigenstates $|\phi\rangle$ at each of the four maximal Wyckoff positions, however each yield one more $C_2$ eigenstate of eigenvalue $+1$ than $-1$.
	Again, all other local eigenstates of the chosen Wyckoff position create an equal number of eigenfunctions of each $\pm$ eigenvalue, so the RSIs are $\delta^2_{1a, 1} = \delta^2_{1b, 1} = \delta^2_{1c, 1} = \delta^2_{1d, 1} = -1$.
	
	These RSI values at each maximal Wyckoff position can be generalized to those in our other line-graph lattices based upon the line-graph graph element sitting on the maximal Wyckoff position and the point-group symmetry; we tabulate these relationships in Table \ref{table:RSI} and prove them in Appendix C of the Supplementary Material \cite{SM}.
	
	\begin{table}[tb]
		\begin{centering}
			\begin{tabular}{l@{\hspace{14pt}}c@{\hspace{4pt}}c@{\hspace{4pt}}c}
				\hline
				&$C_2$& $C_3$& $C_6$\\
				\hline
				vertex&$\delta^2_{w,1}=-1$&$\delta^3_{w,1}=0$&$\delta^6_{1a,1}=\delta^6_{1a,2}=\delta^6_{1a,3}=0$\\
				\multirow{2}{1cm}{complete subgraph} &\multirow{2}{1.5cm}{$\delta^2_{w,1}=+1$}&\multirow{2}{1.5cm}{$\delta^3_{w,1}=+1$}&\multirow{2}{3.7cm}{$\delta^6_{1a,1}=\delta^6_{1a,2}=\delta^6_{1a,3}=+1$}\\\\
				face&$\delta^2_{w,1}=0$&$\delta^3_{w,1}=0$&$\delta^6_{1a,1}=\delta^6_{1a,2}=\delta^6_{1a,3}=0$\\
				\hline
			\end{tabular}
			\caption{For a maximal Wyckoff position $w$ associated with a given point-group symmetry, indicated in the header row, its RSIs can be determined based on the line-graph graph element occupying $w$.}\label{table:RSI}
		\end{centering}
	\end{table}
	
	\section{From RSIs to representation}
	
	Once the RSIs have been determined, it is straightforward to solve for the representation.
	RSIs are linear invariant under induction, so they also describe the differences in EBR multiplicities $\widetilde{m}_{w, k}$ for EBRs induced from the orbitals corresponding to $C_s$ eigenvalue $e^{i2\pi k/s}$ at maximal Wyckoff positions $w$.
	There is also an additional constraint on the total number of flat bands $D$:
	\begin{equation}
	\sum_{w, \,k\in[0,s/2]} m_{w, k} \widetilde{m}_{w, k} = D\label{eq:summ}
	\end{equation}
	where $m_{w, k}$ is the dimension of the induced EBR at maximal Wyckoff position $w$.
	The representations for various families of line-graph lattices are derived in Appendix E of the Supplementary Material \cite{SM}; we now explicitly consider our two examples.
	
	For $C_6$-symmetric lattices we have $m_{1a, 0} = 1$, $m_{2b, 0} = 2$, and $m_{3c, 0} = 3$.
	In example 1, with Eq. (\ref{eq:summ}) we find $\widetilde{m}_{1a, 0} = -1$, $\widetilde{m}_{3c, 0} = 1$ and hence the representation can be written as $(A)_{3c}\!\uparrow \!G \ominus (A)_{1a} \!\uparrow \!G$, where now we use the irrep notation from the Bilbao Crystallographic Server \cite{Elcoro2017}.
	Although this decomposition is not unique, all equivalent decompositions have a negative coefficient.
	Because this representation can be written as a difference of EBRs, the flat bands in example 1---the line graph of the triangle lattice---exhibit fragile topology.
	The Wilson loop for these bands exhibits winding, confirming our result [see Fig. \ref{fig:RSI}(b)].
	
	For $C_2$-symmetric lattices we have $m_{1a, 0} = m_{1b, 0} = m_{1c, 0} = m_{1d, 0} = 1$, so in example 2 we find $\widetilde{m}_{1a, 0} = \widetilde{m}_{1b, 0} = \widetilde{m}_{1c, 0} = \widetilde{m}_{1d, 0} = +1$.
	This yields the representation $(A)_{1a}\!\uparrow \!G \oplus (A)_{1b} \!\uparrow \!G \oplus (A)_{1c} \!\uparrow \!G \oplus (A)_{1d} \!\uparrow \!G$ and we cannot conclude that these four-fold-degenerate bands of example 2 exhibit fragile topology.
	Correspondingly, the Wilson loop eigenvalues show no odd winding.\newline
	
	At this point, among our line-graph lattices we find one $D=2$ lattice with fragile topological flat bands---the line graph of the triangle lattice---and one $D=2$ lattice which admits a Wannier representation---the line graph of the nonagon-triangle lattice (see Appendix F of the Supplementary Material \cite{SM}).
	We also find that all flat-band representations for the $D=3$ and $D=4$ line-graph lattices considered are a sum of EBRs, indicating that each group of bands may be topologically trivial.
	However, we can split the flat-band band degeneracy for these $D>2$ line-graph lattices and characterize the resulting band topology.
	We examine perturbations that leave twofold-degenerate gapped flat bands at the original flat-band energy $-2$.
	We refer to this process as ``splitting the bands''.
	
	\section{Splitting the bands}
	
\begin{figure}[tb]
		\includegraphics[width=\columnwidth]{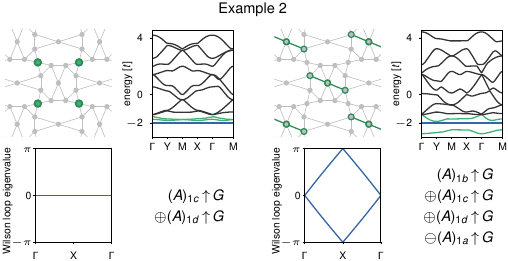}
		\caption{By adding a perturbation (green) consisting of (left) on-site energies or (right) additional hoppings, we can split the $D=4$ degeneracy of example 2 to create 2-fold-degenerate gapped flat bands. We predict the topology of these bands given the perturbation; here the hopping perturbation leaves bands with fragile topology, as seen in the representation and Wilson loop winding.}\label{fig:split}
	\end{figure}

	To begin, we note that on-site-energy perturbations can successfully split the bands for $D=3$ and $D=4$ into flat band(s) and dispersive bands, for example as in the left of Fig. \ref{fig:split}.
	However, the remaining flat band(s) are still EBRs or sums of EBRs.
	Because these perturbations are localized on single vertices, they will not change the existing Wannier representation for the flat-band eigenfunctions.
	
	Therefore, we focus on symmetry-preserving perturbations consisting of new hoppings.
	For $D=4$ line-graph lattices with $C_2$ symmetry, such as example 2 (see Fig. \ref{fig:split}), we find that the bands can always be split into a set of two flat bands and two dispersive ones.
	More specifically, every $D=4$ line-graph lattice has a root-graph unit cell with either two even- and two odd-sided faces (the ``2e2o'' family) or four odd-sided faces (the ``4o'' family).
	For 2e2o lattices, the flat-band degeneracy can be split by introducing a hopping between the two vertices that are each adjacent to both even-sided faces, as shown in Fig. S12(a) \cite{SM}.
	For 4o lattices, it can be split through two hoppings that: (1) are $C_2$ images of one another and share a vertex at a maximal Wyckoff position, (2) each extend across a single face, and (3) are between vertices adjacent to all four faces.
	A construction is depicted in Fig. S13(a) \cite{SM}, with the result seen in the right of Fig. \ref{fig:split}.
	In both families, these prescribed hoppings always exist; of course, there may also be alternate hopping perturbations for these lattices that also split the bands successfully.
	These claims are proved in Appendix D of the Supplementary Material \cite{SM}.
	
	By contrast, for all other line-graph lattices considered we find evidence, presented in Appendix D of the Supplementary Material \cite{SM}, that the bands cannot be split into twofold-degenerate gapped flat bands.
	For example, in $D=3$ lattices with $C_2$ symmetry it seems that hopping perturbations can at best split the three bands into one flat band, sharing a band touch with one dispersive band, and one other, separate, dispersive band.
	
	For bands that can be split, their post-perturbation representation can be predicted with the same formalism.
	Intuitively, a perturbation splits the bands by inducing level repulsion between identical atomic orbitals; indeed, this is the case for example 2 as seen in the right of Fig. \ref{fig:split}, where the perturbed bands each have a representation induced from an orbital on the same maximal Wyckoff positon, $1a$.
	Level repulsion can also occur between two orbitals on general (nonmaximal) Wyckoff positions, which is equivalent to one $s$ and one $p$ orbital for a maximal Wyckoff position $w$ of multiplicity 1 (see Appendix E of the Supplementary Material and the last two rows of Table S3 in \cite{SM}).
	We also find that bands with fragile topology can be realized through our constructed hopping perturbations on the 4o lattices, but not on the 2e2o lattices; proofs are in Appendix E of the Supplementary Material \cite{SM}.
	There we also tabulate representations for perturbed $D=3$ $C_2$-symmetric lattices, where if the perturbation is symmetry-preserving and involves two vertices on a face that sits on a maximal Wyckoff position, then the resulting band pair exhibits fragile topology.\newline
	
	\section{Conclusion}
	
	We have shown how to predict the representation of the energy $= -2$ flat bands for line-graph lattices of planar regular root-graph lattices where these bands are gapped from the rest of the spectrum.
	These predictions only require knowledge of purely geometric qualities of the root-graph lattice structure.
	We further demonstrate that in cases of flat bands with four-fold flat-band degeneracy, perturbations to the line graph always exist to partially break the degeneracy and leave doubly degenerate gapped flat bands, whose representation can also be predicted.
	Of the line-graph lattices considered in this work, we find one $D=2$ lattice with fragile topological flat bands---the line graph of the triangle lattice---and a family of $D=4$ lattices with fragile topological flat bands after one of a class of specific perturbations---the 4o family.
	We also find that for our $D=3$ lattices, there exists a perturbation that yields a pair of fragile topological bands (one flat and one dispersive).
	
	Possible extensions of this work, some of which are briefly discussed in Appendix G of the Supplementary Material \cite{SM}, include extending the formalism to higher degeneracies $D>4$, which will also allow for the treatment of lattices with $C_4$ symmetry, and the addition of $p$- and $d$-orbital hopping to the tight-binding model.
	Other extensions include considering irregular root-graph lattices where vertices can have differing coordination number, nontoroidal root-graph lattices where edges can cross each other without meeting at a vertex, or proving the results of alternate hopping perturbation constructions.
	Similar work has been done on the band topology of ungapped flat bands in line graph and split graphs of bipartite lattices, after the bands are gapped by introducing spin-orbit coupling \cite{Ma}.
	
	Our results dictate the course of quantum simulation of fragile topology in line-graph lattices, a system particularly suitable for the platform of microwave quantum circuits.
	Coplanar waveguide resonators have been used to create various line-graph lattice geometries in two dimensions; in particular, the isotropic three-way capacitor is a well-established and straightforward circuit element to realize such lattices with $d=3$ \cite{Underwood2016, Kollar2019a}.
	By creating artificial materials with these crystalline structures using microwave resonators, it may be possible to probe the physics of fragile topology in flat electronic bands.

	\begin{acknowledgments}
		We acknowledge support from NSF-MRSEC (No. DMR-1420541) and the Princeton Center for Complex Materials, ARO-MURI (No. W911NF-15-1-0397), the NationalKey R\&D Program of China (No. 2016YFA0300600), the National Natural Science Foundation of China (No. 11734003), DOE (No. DE-SC0016239), the Schmidt Fund for Innovative Research, Simons Investigator Grant (No. 404513), the Packard Foundation, NSF-EAGER (No. DMR-1643312), ONR (No. N00014-20-1-2303), the Gordon and Betty Moore Foundation (No. GBMF8685), and US-Israel BSF (No. 2018226).
	\end{acknowledgments}

	\clearpage
	\setcounter{figure}{0}
	\setcounter{table}{0}
	\setcounter{equation}{0}
	\makeatletter
	\renewcommand{\thefigure}{S\@arabic\c@figure}
	\renewcommand{\thetable}{S\@arabic\c@table}
	\makeatother
	\renewcommand{\theequation}{S\arabic{equation}}
    \newcounter{Sfig}
    \renewcommand{\theSfig}{S\arabic{Sfig}}
    \newcounter{Stab}
    \renewcommand{\theStab}{S\arabic{Stab}}
    \renewcommand{\bibnumfmt}[1]{[S#1]}
	
	\appendix
	
	\section{Root- and line-graph lattice properties}\label{appx:rootlineprop}
	
	\begin{figure*}[t]
		\centering
		\begin{minipage}[c]{\textwidth}
			\includegraphics[width=0.95\textwidth]{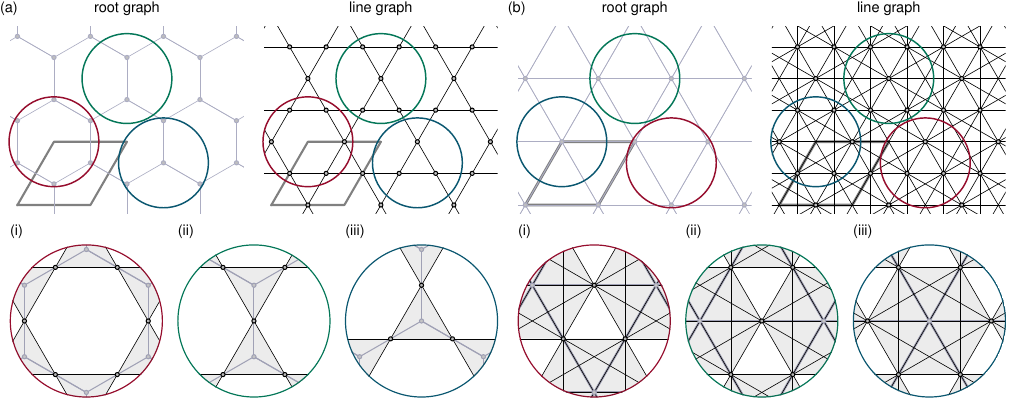}
			\refstepcounter{Sfig}\label{sfig:roottoline}
			\caption{Construction of a line-graph lattice. \textbf{(a)} Starting from a hexagon lattice, we take its line graph and create the kagome lattice. \textbf{(b)} Starting from a triangle lattice, we take its line graph. In both subfigures, \textbf{(i)} highlights how a face of the root graph gives rise to a face of the same number of edges in the line graph; \textbf{(ii)} highlights how an edge of the root graph gives rise to a vertex in the line graph; and \textbf{(iii)} highlights how a vertex of the root graph gives rise to a complete subgraph of $d$ vertices, where $d$ is the coordination number of the root-graph vertex. In the line-graph lattice, we refer to complete subgraphs only as complete subgraphs and not by triangles or faces, to draw a distinction from the faces in the line graph which originate from faces in the root graph. The parallelograms outlined in grey denote a single unit cell of each lattice.}
		\end{minipage}
	\end{figure*}
	
	Here we prove the relevant properties of line graphs and line-graph lattices used in the main text. As a reminder, we consider regular root-graph lattices $X$ that can be embedded on a torus and have coordination number $d$ and $n$ vertices per unit cell, and their associated line-graph lattices $L(X)$.
	\begin{enumerate}
		\item[LG1] \emph{If $X$ is a periodic lattice, $L(X)$ is as well.}
		\item[LG2] \emph{Any symmetries of $X$ are inherited by $L(X)$, \textit{i.e.} the space group of $X$ is the same as that of $L(X)$.}
	\end{enumerate}
	These two properties follow from the fact that to define $X$ as a lattice, we embed its vertices in the Euclidean plane where distances are well-defined. Because the vertices of $L(X)$ can be defined completely in terms of the positions of the vertices of $X$, all periodicity and symmetry relations of $X$ are inherited by $L(X)$.
	\begin{enumerate}
		\item[LG3] \emph{As a consequence of the line-graph construction, every vertex $v_{X, i}$ of the root graph gives rise to a ``complete subgraph'' in the line graph, where a complete subgraph is defined as a subset of $k$ vertices and binomial coefficient $\binom{k}{2}$ edges for which all pairs of vertices are connected by one of the edges. In these complete subgraphs, $k$ will be equal to the coordination number of $v_{X,i}$.}
	\end{enumerate}
	In Figure \ref{sfig:roottoline} we show two examples of line-graph lattice constructions as a reference.
	In subfigure (a), we show the line graph of the hexagon lattice; although this is biparitite and will give rise to an ungapped flat band, we include it for its simplicity.
	In subfigure (b), we show the line graph of the triangle lattice (Example 1 of the main text) for its relevance to this work.
	
	Consider all edges $(e_{X, 1}, e_{X, 2}, \dots e_{X, d})$ of the root graph $X$ that have an endpoint at a vertex $v_{X, i}$; there are $d$ of these edges because the root graph has coordination number $d$.
	Upon taking the line graph, see (iii) of Figure \ref{sfig:roottoline}(a) and (b), these root-graph edges will give rise to line-graph vertices $(v_{L(X), 1}, v_{L(X), 2}, \dots v_{L(X), d})$.
	All of these vertices will be connected to one another by an edge, because the root-graph edges they originate from all share a vertex ($v_{X, i}$).
	This set of vertices and edges of the line graph then make up the complete subgraph defined in the \ref{itm:LGsubgraph} property statement, with $k=d$.
	
	For clarity, we will refer to these complete subgraphs only as complete subgraphs and not by any of their vertices, edges, or faces.
	In particular, although the complete subgraph arising from a root-graph vertex of coordination number $3$ looks like a triangle, we will refer to it as a complete subgraph rather than as a triangle or face.
	This is to avoid confusion with the line-graph faces that arise from faces in the root graph as discussed in \ref{itm:LGcycle}.
	
	\begin{enumerate}
		\item[LG4] \emph{Consider a sequence of vertices of the root graph $(v_{X, 1}, v_{X, 2}, \dots v_{X, n+1})$ where $v_{X, 1}=v_{X, n+1}$ but all other vertices are distinct. Take the sequence of edges $(e_{X, 1}, e_{X, 2}, \dots e_{X, n})$ of $X$ where the edge $e_{X, i}$ connects vertices $v_{X, i}$ and $v_{X, i+1}$. These vertices and edges form a ``cycle'' of the graph. As a consequence of the line-graph construction, every cycle of $X$ gives rise to a cycle of equal length (number of edges) of $L(X)$. These cycles of $L(X)$ are ``chordless'', meaning that no two vertices of the cycle are connected by an edge that does not belong in the cycle.}
	\end{enumerate}
	Take the line graph of $X$ as in Figure \ref{sfig:roottoline}; then as in (i) of subfigures (a) and (b), the root-graph cycle's edges give rise to a sequence of vertices $(v_{L(X), 1}, v_{L(X), 2}, \dots v_{L(X), n})$ of the line-graph lattice.
	Vertices $v_{L(X), i}$ and $v_{L(X), i+1 \mod n}$ will be connected by an edge, because the edges they come from join a sequence of vertices.
	This set of vertices and edges of the line graph then make up a cycle of length equal to that of the root-graph cycle considered, $n$.
	Moreover, this line-graph cycle must be chordless; for two vertices $v_{L(X), j}$ and $v_{L(X), k}$ of the line-graph cycle to be connected by an edge, we must have that the root-graph edges $e_{X, j}$ and $e_{X, k}$, which gave rise to those line-graph vertices, share a vertex.
	Because all vertices of the root-graph cycle must be distinct, the edges $e_{X, j}$ and $e_{X, k}$ can only share a vertex if $j$ and $k$ differ by $1$ (mod $n$), \textit{i.e.} if the edge connecting $v_{L(X), j}$ and $v_{L(X), k}$ is part of the cycle.

	As a result, the cycles bounding each face of the root-graph lattice give rise to cycles of corresponding equal length in the line-graph lattice.
	We will refer to these latter cycles specifically as ``faces'' of the line-graph lattice, separate from the complete subgraphs of \ref{itm:LGsubgraph}.
	
	\begin{enumerate}
		\item[LG5] \emph{Given energies $E_X$ of $X$, its corresponding line-graph lattice $L(X)$ has energies $E_{L(X)} = \{E_X + d - 2, -2\}$, with one or more flat bands at $-2$.}
	\end{enumerate}
	The existence of least eigenvalue $-2$ for general line graphs of regular graphs is well-known among the graph theory community \cite{Cvetkovic2004s}; a proof extending this to lattices can be found in \cite{Kollar2019}.
	\begin{enumerate}
		\item[LG6] \emph{The degeneracy $D$ of the flat band at $-2$ is given by $D = n(d-2)/2$.}
	\end{enumerate}
	The root-graph lattice has $n$ vertices per unit cell and therefore $n$ bands.
	The line-graph lattice has a number of vertices per unit cell equal to the number of edges in the root-graph lattice; this can be straightforwardly counted to be $nd/2$.
	As a result, the line-graph lattice has $nd/2$ electronic bands.
	Of these bands, $n$ correspond directly to the $n$ bands of the root-graph lattice, shifted in energy by $d-2$ as noted in property \ref{itm:LGspectrum}.
	The remaining bands must be flat and at energy $-2$; there are $D = nd/2-n = n(d-2)/2$ of these bands.
	\begin{enumerate}
		\item[LG7] \emph{If $X$ is non-bipartite, then the flat band(s) at $-2$ for $L(X)$ will be gapped from the other bands.}
	\end{enumerate}
	The proof for this can be found in \cite{Kollar2019}.
	\begin{enumerate}
		\item[LG8] \emph{The number of faces per unit cell of $X$ is also given by $n(d-2)/2$.}
	\end{enumerate}
	Euler's formula for graphs that can be embedded on a torus (without edge crossings) states that the number of faces is given by the difference in the number of edges and vertices.
	Considering a single unit cell, the number of faces in the root-graph lattice is then $nd/2 - n = n(d-2)/2$.
	
	Given that we must have at least $3$ edges per face, the integer solutions for $(n, d)$ given $D$ are:
	
	\noindent $\mathbf{D=2}$: $(n, d) = (4, 3), (2, 4), (1, 6)$ 
	
	\noindent $\mathbf{D=3}$: $(n, d) = (6, 3), (3, 4), (2, 5)$ 
	
	\noindent $\mathbf{D=4}$: $(n, d) = (8, 3), (4, 4)$ 
	
	For a given $(n, d)$ pair, we can additionally tabulate the possible number of edges per face for the faces in the root-graph unit cell, under the constraint that there must be faces with an odd number of edges to keep the root graph non-bipartite.
	The results are found in Table \ref{stab:geoms}.
	
	\begin{figure*}[ht]
		\begin{minipage}[b]{0.49\linewidth}
			\centering
			\includegraphics[width=\columnwidth]{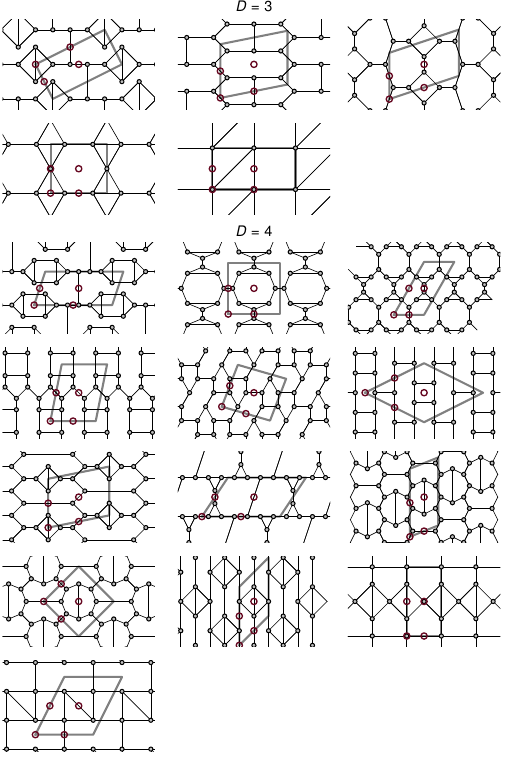}
			\refstepcounter{Sfig}\label{sfig:C2}
			\caption{Examples of root-graph lattices with $C_2$ symmetry and $1 < D \leq 4$. For each lattice, one unit cell is outlined in grey and maximal Wyckoff positions are circled in red.}
		\end{minipage}
		\hspace{0.5cm}
		\setcounter{figure}{0}
		\renewcommand{\figurename}{TABLE}
		\begin{minipage}[b]{0.45\linewidth}
			\centering
			\begin{tabular}{l@{\hspace{12pt}}l@{\hspace{24pt}}l}
				\hline
				&$(n, d)$&numbers of edges in each face\\
				\hline
				$D=2$&$(4, 3)$&9, 3\\
				&$(2, 4)$&-\\
				&$(6, 1)$&3, 3\\
				$D=3$&$(6, 3)$&12, 3, 3*\\
				&&8, 5, 5\\
				&&4, 7, 7\\
				&$(3, 4)$&6, 3, 3*\\
				&$(2, 5)$&4, 3, 3\\
				$D=4$&$(8, 3)$&14, 4, 3, 3\\
				&&12, 6, 3, 3\\
				&&10, 8, 3, 3\\
				&&10, 4, 5, 5\\
				&&8, 6, 5, 5\\
				&&6, 4, 7, 7\\
				&&9, 9, 3, 3*\\
				&&7, 7, 5, 5*\\
				&&7, 7, 7, 3\\
				&$(4, 4)$&5, 5, 3, 3*\\
				&&6, 4, 3, 3\\
				\hline
			\end{tabular}
			\refstepcounter{Stab}\label{stab:geoms}
			\caption{Possible numbers of edges in each face of the root-graph unit cell, for the toroidal non-bipartite root-graph lattices considered in this work.
				Given flat-band degeneracy $D$ of the line-graph lattice, several pairs of coordination numbers $d$ and number of vertices per unit cell $n$ for  the root-graph lattice are possible.
				Then for each $(n, d)$ pair, the possible number of edges in each face of the root-graph unit cell can be determined.
				For all of these entries, we have verified that at least one root-graph lattice geometry exists, see Figures \ref{sfig:C2} and \ref{sfig:C3C6}.
				For entries with an asterisk, we have verified that at least two root-graph lattice geometries exist.}
		\end{minipage}
		\setcounter{figure}{2}
		\setcounter{table}{1}
	\end{figure*}

	\section{Real-space eigenfunctions}\label{appx:eigenfn}
	
\begin{figure*}[t]
		\centering
		\begin{minipage}[c]{\textwidth}
			\includegraphics[width=\textwidth]{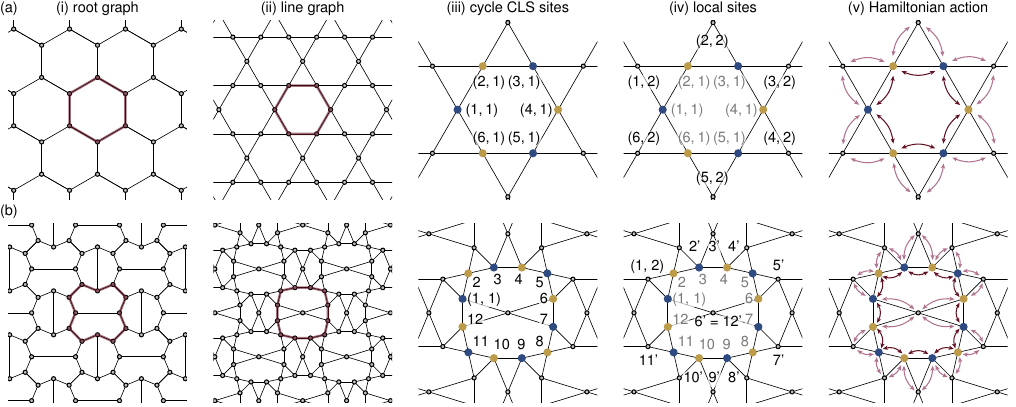}
			\refstepcounter{Sfig}\label{sfig:cycleCLS}
			\caption{Construction of a cycle CLS. \textbf{(a)} Hexagon root-graph lattice. \textbf{(b)} Example 2 of the main text. In both examples, we start in \textbf{(i)} by identifying a cycle of even number of edges, which gives rise in \textbf{(ii)} to a chordless cycle of the same number of edges in the line-graph lattice. Notice that in (b)(ii), the vertex enclosed by the even cycle is not part of the cycle, so the cycle is indeed chordless. In \textbf{(iii)}, we show the labeling for vertices that are part of the cycle and where the real-valued wavefunction has non-zero amplitude. Blue (yellow) circles denote relative wavefunction amplitude $+1$ ($-1$). In (b), labels $(x, 1)$ have been abbreviated as $x$ for legibility. We then label neighboring vertices that are not part of the cycle and have zero wavefunction amplitude in \textbf{(iv)}, where in (b) labels $(x, 2)$ have been abbreviated as $x'$, to determine the action of the tight-binding Hamiltonian \textbf{(v)}. Red arrows denote hopping between two vertices where the wavefunction has nonzero amplitude; pink arrows denote hopping between two vertices where the wavefunction has nonzero amplitude on one and zero amplitude on the other. Notice that these cycle CLS energy eigenstates can be described by the faces that they encircle.}
		\end{minipage}
	\end{figure*}

	In this appendix is a construction of several classes of energy $-2$ eigenstates that are generally non-orthonormal but span the flat-band basis: cycle and chain compact localized states (CLSes) \cite{Sutherland1986, Flach2014}, as well as extended states.
	The set of all cycle CLSes, chain CLSes, and extended states forms an overcomplete basis, however we will show that a subset of these states can be chosen to construct a complete basis.
	Cycle CLSes can be constructed in the following way \cite{Kollar2019}, depicted in Figure \ref{sfig:cycleCLS}: to begin, if a regular root graph of degree $d$ has a cycle of even length (\emph{e.g.} Figure \ref{sfig:cycleCLS}(i), highlighted in red), this length can be written as $2k$ where $k$ is a positive integer.
	Upon taking the line graph operation, as described in property \ref{itm:LGcycle} of line graphs, vertices are created on the $2k$ edges of this even-length cycle and connected in a chordless cycle of equal length (Figure \ref{sfig:cycleCLS}(ii), highlighted in red).
	Now label the vertices of this new cycle as $(x, 1)$ where $x \in [1, 2k]$ denotes a vertex numbering within the cycle such that vertex $(x, 1)$ is connected to vertices $(x\pm1 \mod 2k, 1)$ (Figure \ref{sfig:cycleCLS}(iii)).
	Define the cycle CLS to be the wavefunction
	\begin{align}
	\psi((x,i))=
	\begin{cases}
	1 &x \leq 2k, x \ \mathrm{even}, i=1\\
	-1 &x \leq 2k, x \ \mathrm{odd}, i=1\\
	0 &\mathrm{i\neq1}
	\end{cases}\label{eq:cycleCLS}
	\end{align}
	
	To evaluate how the Hamiltonian $\hat{H}_{L(X)} = \sum_{\langle j, l\rangle} \hat{a_j}^\dagger \hat{a_l} + \hat{a_l}^\dagger \hat{a_j}$, where $j$ and $l$ are vertices connected by edges (adjacent), acts on the cycle CLS, we also label all additional vertices adjacent to those in the even cycle under consideration of the line-graph lattice (Figure \ref{sfig:cycleCLS}(iv)).
	As shown in Figure \ref{sfig:cycleCLS}, these additional vertices are in complete subgraphs and adjacent to two vertices in the even cycle; thus, for vertices not within the cycle but within a complete subgraph sharing vertices $(x, 1)$ and $(x+1 \mod 2k, 1)$ with the cycle, label them as $(x, i \in [2, d-1])$.
	While this labeling is neither unique nor complete, and may assign multiple labels to a vertex, it is sufficient for self-consistently defining an eigenstate.
	
	Then upon applying the Hamiltonian $\hat{H}_{L(X)}$, we find
	
	\begin{widetext}
		\begin{align}
		\hat{H}_{L(X)}\psi((x,i))&=
		\begin{cases}
		\!\begin{aligned}[b]
		& \sum_{i'=1}^{d-1} \psi((x-1 \mod 2k, i')) + \sum_{i'=2}^{d-1} \psi((x, i')) + \psi((x+1 \mod 2k, i))
		\end{aligned}  &x \leq 2k, i=1\\
		\!\begin{aligned}[b]
		&\sum_{i'=1}^{d-1}\psi((x, i')) - \psi((x, i)) + \psi((x+1 \mod 2k, 1))
		\end{aligned} &x \leq 2k, i\neq 1\\
		0 &\mathrm{else}
		\end{cases}\\
		&=-2\psi((x,i))
		\end{align}
	\end{widetext}
	as shown schematically in Figure \ref{sfig:cycleCLS}(v).
	Therefore, we find that for every even-length cycle of the root graph, which gives rise to an even cycle of the line graph, there exists one cycle CLS flat-band energy eigenstate for the line graph, which has nonzero amplitude on the vertices of the line-graph even cycle.
	Because these root-graph cycles bound one or more faces of the root graph, we will often refer to these cycle CLSes of the line graph by these faces that they encircle.
	For example, in Figure \ref{sfig:cycleCLS} we would refer to the cycle CLS of subfigure (a)(iii) as ``encircling a hexagon'', and the cycle CLS of (b)(iii) as ``encircling two heptagons'', based on the root-graph faces bounded by the root-graph cycles from which these CLSes originate.
	
\begin{figure*}[t]
		\centering
		\begin{minipage}[c]{\textwidth}
			\includegraphics[width=\textwidth]{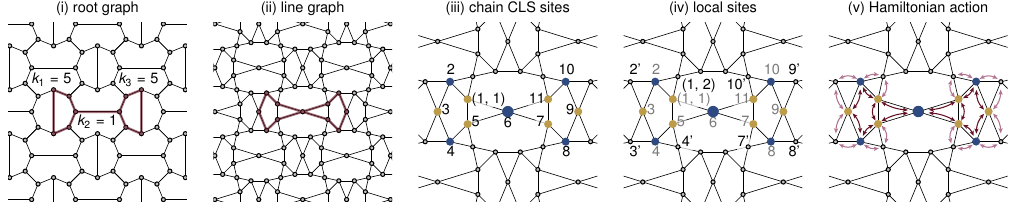}
			\refstepcounter{Sfig}\label{sfig:chainCLS}
			\caption{Construction of a chain CLS. Because the construction relies on the existence of odd-sided faces, there are no chain CLSes in the line graphs of bipartite root-graphs. Hence we do not show the hexagon/kagome lattice example here and show only Example 2 from the main text. \textbf{(i)} We begin by identifying two cycles in the root graph of odd number of edges that do not share vertices (of lengths $k_1=5$ and $k_3=5$, plus a path (of length $k_2=1$) between a vertex in one and a vertex in the other. From the corresponding graph elements that are created when taking the line graph \textbf{(ii)}, a real-valued wavefunction can be constructed and vertices with non-zero amplitude labeled \textbf{(iii)}. Blue (yellow) circles denote positive (negative) relative wavefunction amplitude, and the larger circle at vertex $6$ has twice the amplitude of the remaining, smaller circles. \textbf{(iv)} Neighboring vertices are also labeled to determine the action of the Hamiltonian on the wavefunction.  Here, labels $(x, 1)$ have been abbreviated as $x$ and labels $(x, 2)$ have been abbreviated as $x'$ for legibility.  \textbf{(v)} Red arrows denote hopping between two vertices where the wavefunction has nonzero amplitude; pink arrows denote hopping between two vertices where the wavefunction has nonzero amplitude on one and zero amplitude on the other. Notice that chain CLSes, like cycle CLSes, can be described by the faces that they encircle.}
		\end{minipage}
	\end{figure*}

	As seen in Figure \ref{sfig:chainCLS}, we can also construct line-graph energy $-2$ eigenstates as chain CLSes.
	Begin with two odd cycles that do not share any vertices (are non-adjacent) of lengths $k_1$ and $k_3$ in the root graph, connected by a path of length $k_2>0$ edges, as shown in Figure \ref{sfig:chainCLS}(i).
	In this example, $k_1=k_3=5$ and $k_2=1$.
	Then as in Figure \ref{sfig:chainCLS}(ii) the line graph has two corresponding even cycles of lengths $k_1+1$ and $k_3+1$, connected by a path of $k_2$ vertices.
	Beginning and ending where the cycles meet the vertices or vertex of the connecting path (see Figure \ref{sfig:chainCLS}(iii) and (iv)), label the vertices $(x, i)$ as for the cycle CLS in Figure \ref{sfig:cycleCLS}.
	Here $x\in [1, k_1+k_2+k_3]$ and $i\in[1,d-2]$ if $x=k_1, k_1+k_2$, or $k_1+k_2+k_3$, otherwise $i \in [1, d-1]$.
	
	Then we have
	
	\begin{align}
	\psi((x,i))=
	\begin{cases}
	1 &x \leq k_1, x \ \mathrm{even}, i=1\\
	-1 &x \leq k_1, x \ \mathrm{odd}, i=1\\
	2 &k_1 < x \leq k_1+k_2, x \ \mathrm{even}, i=1\\
	-2 &k_1 < x \leq k_1+k_2, x \ \mathrm{odd}, i=1\\
	1 &k_1+k_2<x\leq k_1+k_2+k_3, x \ \mathrm{even}, i=1\\
	-1 &k_1+k_2<x\leq k_1+k_2+k_3, x \ \mathrm{odd}, i=1\\
	0 &\mathrm{else}
	\end{cases}
	\end{align}
	It is straightforward to confirm that this state, too, is an energy $-2$ eigenstate (see Figure \ref{sfig:chainCLS}(v)).
	In analogy with cycle CLSes, we will also often refer to chain CLSes by the two sets of faces that they encircle, one at each end.
	
\begin{figure*}[t]
		\centering
		\begin{minipage}[c]{\textwidth}
			\includegraphics[width=\textwidth]{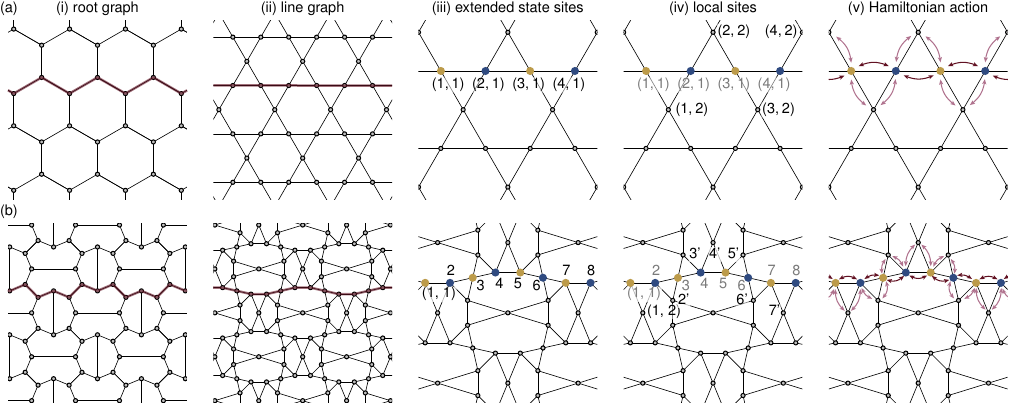}
			\refstepcounter{Sfig}\label{sfig:extended}
			\caption{Construction of an extended state. \textbf{(a)} Hexagon root-graph lattice. \textbf{(b)} Example 2 of the main text. In both examples, we start in \textbf{(i)} by identifying a cycle of an even number of edges that wraps around the toroidal or poloidal direction of the torus. \textbf{(ii)} This gives rise to a similar cycle in the line graph lattice, whose vertices \textbf{(iii)} can be labeled \textbf{(iv)} along with its neighboring vertices to determine \textbf{(v)} the action of the Hamiltonian on a wavefunction with nonzero amplitude on those vertices. Blue (yellow) circles denote relative wavefunction amplitude $+1$ ($-1$). Red arrows denote hopping between two vertices where the wavefunction has nonzero amplitude; pink arrows denote hopping between two vertices where the wavefunction has nonzero amplitude on one and zero amplitude on the other. In (b), labels $(x, 1)$ have been abbreviated as $x$ and labels $(x, 2)$ have been abbreviated as $x'$ for legibility.}
		\end{minipage}
	\end{figure*}

	Finally, an extended state, or noncontractible loop state, consists of a wavefunction with nonzero amplitude on a chordless, even cycle around the torus where the amplitudes are real-valued, equal in amplitude, and alternate in sign.
	As shown in Figure \ref{sfig:extended}, the vertices on and adjacent to the cycle can be labeled in the same way as for cycle CLSes, with extended state wavefunction given by Equation \ref{eq:cycleCLS}.
	They are also energy $-2$ eigenstates.
	
\begin{figure*}[t]
		\centering
		\begin{minipage}[c]{\textwidth}
			\includegraphics[width=\textwidth]{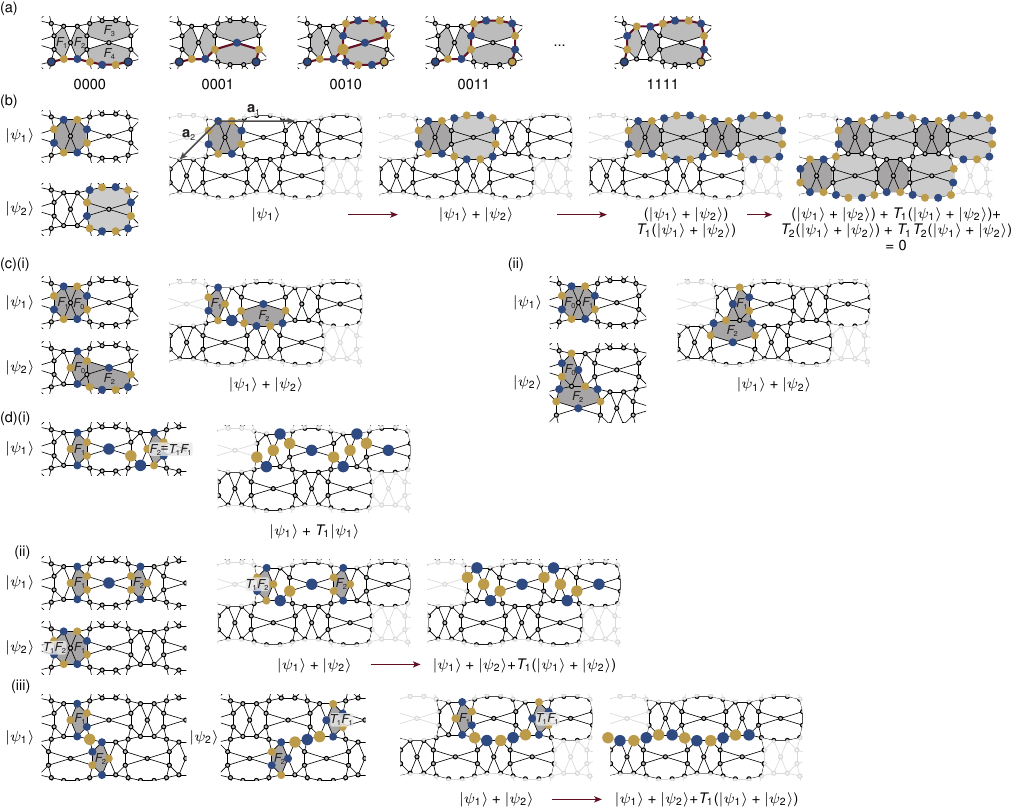}
			\refstepcounter{Sfig}\label{sfig:basisfacts}
			\caption{Depiction of flat-band-state properties presented in Appendix \ref{appx:eigenfn}, useful for constructing flat-band bases. Here, we show a $2$-unit-cell $\times 2$-unit cell line graph of the heptagon-heptagon-pentagon-pentagon lattice with $M_x$ and $M_y$ mirror symmetries (\emph{i.e.} Example 2 of the main text), with periodic boundary conditions. In all subfigures, blue (yellow) circles denote positive (negative) relative wavefunction amplitude, and larger circles have twice the amplitude of the smaller circles.
			\textbf{(a)} Encoding of extended states into binary strings of length $D$, for the purpose of determining the number of extended states that are linearly independent from a given set of cycle and chain CLSes. See text for details.
			\textbf{(b)} Schematic for \ref{itm:FBannihil}, showing how a set of  CLSes that cover the torus exactly must be linearly dependent, because adding them together yields the zero function. Here and in the remaining subfigures, $\mathbf{a}_1$ and $\mathbf{a}_2$ are the lattice vectors, and $T_1$ ($T_2$) denotes translation by $\mathbf{a}_1$ ($\mathbf{a}_2$).
			\textbf{(c)} Schematic for \ref{itm:FBCLSgen}, showing how a (i) chain and (ii) cycle CLS may be created through linear combinations of cycle CLSes.
			\textbf{(d)(i)}-\textbf{(iii)} Schematics for \ref{itm:FBextgen}, showing how a chain CLS can be combined with chain and cycle CLSes to generate an extended state.}
		\end{minipage}
	\end{figure*}

	We can define as many extended states as there exist even cycles around the torus in the root-graph lattice.
    To enumerate them, we can think of the extended states on the torus in the following way: consider a vertex $v_1$ and its copy $T_1 v_1$, translated by a lattice vector $\mathbf{a}_1$.
	For example, consider a single unit cell of the line-graph lattice, such as the one shown in Figure \ref{sfig:basisfacts}(a) (for a single unit cell, the vertices $v_1$ and $T_1 v_1$ are the same).
	An extended state whose cycle includes these vertices then partitions this unit cell (and the torus) relative to some defined reference partition; we can then examine in which part each face of the unit cell lies, correspondingly encoding the extended state as a binary string as shown in Figure \ref{sfig:basisfacts}(a).
	This binary string has length equal to the number of faces in the unit cell, $D$, comprised of a number of leading bits equal to the number of even-sided faces per unit cell, and a number of trailing bits equal to the number of odd-sided faces per unit cell.
	
	In the example of Figure \ref{sfig:basisfacts}(a), we have $D=4$ bits in the string, comprised of $0$ leading bits and $4$ trailing bits.
	A binary string of $0000$ can be chosen to define our reference, corresponding to an extended state where all four faces are ``above'' the cycle in the depiction of the lattice on a plane; $0001$ then corresponds to the state where three of the faces are ``above'' the cycle and one is ``below''; and so on.
	More generally, for lattices with $M_1 \times M_2$ unit cells, the extended state will consist of the same wavefunction amplitudes repeated across the $M_1$ unit cells along lattice vector $\mathbf{a}_1$, and there will be $M_2$ such extended states, corresponding to repeated lattice translations $T_2$ of the entire extended state by the other lattice vector $\mathbf{a}_2$.
	We note that there may be redundancy of encodings of these extended states; for example, in Figure \ref{sfig:basisfacts}(a) the $1111$ extended state is equivalent to the $0000$ state because under periodic boundary conditions, the vertices on the left and right (top and bottom) edges of the unit cell are the same.
	
	With this limitation in mind, we will use this binary string representation for extended states to find extended states which are linearly independent from a given set of cycle and chain CLSes to construct complete flat-band bases for our flat bands.
	Our use of these extended states should not be mistaken as the extended states being absolutely necessary to describe the flat-band states, as would be a consequence of fragile topological bands.
		
	The existence of these compact localized and extended flat-band eigenstates crucially depends upon the following: for even cycles in the line-graph lattice that are created from even cycles in the root graph, vertices not in the cycle but nearest neighbors with a vertex $v_{L(X), i}$ in the cycle, are also nearest neighbors with a nearest neighbor of $v_{L(X), i}$ that is also in the cycle (see Figures \ref{sfig:cycleCLS}-\ref{sfig:extended}).
	As an example, take the cycle of Figure \ref{sfig:cycleCLS}(b)(ii) and vertex $(1, 1)$ in this cycle with neighbors $2$ and $12$ also in the cycle.
	Vertices not in this cycle, but which are nearest neighbors with $(1, 1)$, are also nearest neighbors with vertices $2$ or $12$.
	To prove this statement, consider a vertex $v_{L(X), i}$ in the cycle of the line graph; it originates from an edge $e_{X, i}$ of the root graph.
	Every edge $e_{X, i}$ of the root graph has a vertex at either end, $v_{X, i}$ and $v_{X, i+1}$, each of which gives rise to a complete subgraph in the line graph.
	In the line graph construction, $v_{L(X), i}$ is then only connected to vertices in these complete subgraphs.
	Indeed, in our example, $(1, 1)$ is part of two complete subgraphs of $k=3$ vertices each (seen as triangles in Figure \ref{sfig:cycleCLS}(b)(iii)).
	Furthermore, the two neighbors of $v_{L(X), i}$ in the cycle are each in one of these two complete subgraphs.
	In our example, the vertex labeled $2$ is in one of the two complete subgraphs, while the vertex labeled $12$ is in the other.
	Then if a vertex $v_{L(X), \alpha}$ is not in the cycle but is nearest neighbors with a vertex $v_{L(X), i}$ in the cycle, it must be in one of the complete subgraphs and therefore a nearest neighbor with one of the two neighbors of $v_{L(X), i}$ in the cycle.
	For example, vertex $(1, 2)$ of Figure \ref{sfig:cycleCLS} is adjacent to $(1, 1)$ but not in the cycle, and we see that it is also adjacent to vertex $2$ of the cycle.

	Upon applying the tight-binding Hamiltonian, this geometry ensures destructive interference of the flat-band eigenfunction's opposite-valued amplitudes on neighboring vertices, on neighboring vertices where the eigenfunction has zero amplitude.
	A similar geometry ensures this destructive interference for chain CLSes, which we do not elaborate upon here.
	We refer to these geometries as ``providing compact support''.

	Notice that under our constructions for cycle and chain CLSes, all even-sided faces of the root-graph lattice contribute a cycle CLS (as in Figure \ref{sfig:cycleCLS}(a)), and any two odd-sided faces of the root-graph unit cell contribute a cycle or chain CLS depending on whether or not they are adjacent (as in Figures \ref{sfig:cycleCLS}(b) and \ref{sfig:chainCLS}).
	Because of this, we often refer to wavefunctions based on the root-graph faces they descend from.
	By contrast, as seen in Figure \ref{sfig:extended}, extended states cannot be described based on the lattice faces, but can be labeled as traveling in the toroidal or poloidal direction of the torus (\emph{i.e.} one of the two lattice vector directions).
	Although in general these states are not orthogonal, a linearly independent and complete basis set can be chosen and proven to be as such based on their parent faces \cite{Bergman2008, Kollar2019}.

	We will define such a basis by beginning with a set of cycle and chain CLSes, removing any linearly dependent states, then adding in all extended states which are linearly independent.
	Before elaborating on this construction, we explore several flat-band-eigenstate linear dependencies, with examples shown in Figure \ref{sfig:basisfacts}, which will allow us to show that our constructed set of basis states are indeed linearly independent and span the flat-band basis.
	This construction will be necessary for Appendix \ref{appx:perturb}, where we add perturbations to split 3- and 4-fold-degenerate flat bands into doubly degenerate flat bands and dispersive bands.
	
	\begin{enumerate}
		\item[\namedlabel{itm:FBannihil}{FB1}] \emph{A linear combination of (equally weighted) CLSes that cover the torus exactly will annihilate.}
	\end{enumerate}
	This is shown in \cite{Kollar2019}.
	Additionally, we provide a schematic depiction in Figure \ref{sfig:basisfacts}(b) for a $2$-unit-cell by $2$-unit-cell lattice with periodic boundary conditions, and a brief summary.
	Given a CLS $|\psi_1\rangle$, we can identify its parent faces and say that the CLS ``covers'' these faces.
	In Figure \ref{sfig:basisfacts}(b), we show this by shading the parent faces in addition to showing the CLS wavefunction amplitudes.
	Then, by adding a second CLS $|\psi_2\rangle$ (see Figure \ref{sfig:basisfacts}(b)), the linear combination covers additional faces. 
	If CLSes are added across all unit cells in such a way that all faces of the lattice are covered exactly once, and the faces covered by any CLS share at most one edge with the faces covered by any other CLS, then the resulting wavefunction will appear to have non-zero amplitude only at the boundaries of the depicted lattice.
	For example, see the last lattice in the series of Figure \ref{sfig:basisfacts}(b), where $T_1$ ($T_2$) is translation by the lattice vector $\mathbf{a}_1$ ($\mathbf{a}_2$).
	However, due to periodic boundary conditions, the vertices drawn at the left and right (top and bottom) of the depicted lattice refer to the same vertices, and the wavefunction is in fact identically zero.
	We note that in this treatment, whether or not complete subgraphs are enclosed by the CLSes does not affect the wavefunction annihilation.
	
	\begin{enumerate}
		\item[\namedlabel{itm:FBCLSgen}{FB2}] \emph{Chain and cycle CLSes can be generated through combinations of cycle CLSes.}
	\end{enumerate}
	Consider two unique cycle CLSes $|\psi_1\rangle$ and $|\psi_2\rangle$ that both encircle an odd-sided face $F_0$, but otherwise encircle unique sets of odd-sided faces $F_1$ and $F_2$, as drawn in Figure \ref{sfig:basisfacts}(c).
	Then the sum $|\psi_1\rangle + |\psi_2\rangle$ will not encircle $F_0$, but will encircle both $F_1$ and $F_2$.
	If there is a vertex at the boundaries of both $F_1$ and  $F_2$, the resulting CLS will be a cycle CLS, as in (c)(ii); otherwise, it will be a chain CLS, as in (c)(i).
	We note that if $F_0$, $F_1$, and $F_2$ are even-sided, then the sum $|\psi_1\rangle + |\psi_2\rangle$ will result in a sum of two cycle CLSes.
	
	\begin{enumerate}
		\item[\namedlabel{itm:FBextgen}{FB3}] \emph{Extended states can be generated through linear combinations of chain and cycle CLSes.}
	\end{enumerate}
	Define a chain CLS $|\psi_1\rangle$ that encircles odd-sided faces $F_1$ and $F_2$.
	For example, $|\psi_1\rangle$ in Figure \ref{sfig:basisfacts}(d)(i) shows a chain CLS that encircles a pentagon on the left, $F_1$, and a pentagon on the right, $F_2$, both shaded in grey.
	Now consider $T_1 |\psi_1\rangle$, where $T_1$ denotes translation by lattice vector $\mathbf{a}_1$; its encircled faces are then $T_1 F_1$ and $T_1 F_2$.
	If $T_1 F_1 = F_2$ (or $T_1 F_2 = F_1$), as in Figure \ref{sfig:basisfacts}(d)(i), then the sum $|\psi_1\rangle + T_1 |\psi_1\rangle$ is itself a chain CLS with encircled faces $F_1$ and $T_1 F_2$ ($F_2$ and $T_1 F_1$).
	Then upon adding these translated copies around the entire torus, we find an extended state: $|\psi_\mathrm{ext}\rangle = \sum_i^{N_1} (T_1)^i |\psi_1\rangle$, where $N_1$ is the number of unit cells around the torus in the $\mathbf{a}_1$ direction (in Figure \ref{sfig:basisfacts}(d)(i), $N_1=2$).
	
	If this is not the case, a second possibility (shown in Figure \ref{sfig:basisfacts}(d)(ii)) is that $T_1 F_1$ and $F_2$ (or $T_1 F_2$ and $F_1$) are \emph{adjacent}, that is, there is a shared vertex at the boundaries of both faces.
	Then a cycle CLS $|\psi_2\rangle$ can be defined to encircle both $T_1 F_1$ and $F_2$ ($T_1 F_2$ and $F_1$); in Figure \ref{sfig:basisfacts}(d)(ii), this is the cycle CLS encircling both pentagons.
	We can then construct the extended state $|\psi_\mathrm{ext}\rangle = \sum_i^{N_1} (T_1)^i (|\psi_1\rangle + |\psi_2\rangle)$.
	
	A third possibility is that neither $T_1 F_1$ and $F_2$, nor $T_1 F_2$ and $F_1$, are adjacent, seen in Figure \ref{sfig:basisfacts}(d)(iii).
	In this case, a second chain CLS $|\psi_2\rangle$ can be defined to encircle faces $F_2$ and $T_1 F_1$, such that $|\psi_1\rangle + |\psi_2\rangle$ is a chain CLS encircling faces $F_1$ and $T_1 F_1$.
	Then we have the extended state $|\psi_\mathrm{ext}\rangle = \sum_i^{N_1} (T_1)^i (|\psi_1\rangle + |\psi_2\rangle)$.
	
	Two quick remarks are in order.
	First, because this extended state generation depends on the existence of chain CLSes, it only applies to line-graph lattices whose root graphs contain faces with an odd number of edges (as required in the construction of chain CLSes).
	Second, the other lattice translation vector $T_2$ by the other lattice vector $\mathbf{a}_2$ can be used in place of $T_1$; in this discussion we have considered $T_1$ without loss of generalization.\newline
	
	\begin{figure}[t]
		\centering
		\includegraphics[width=\columnwidth]{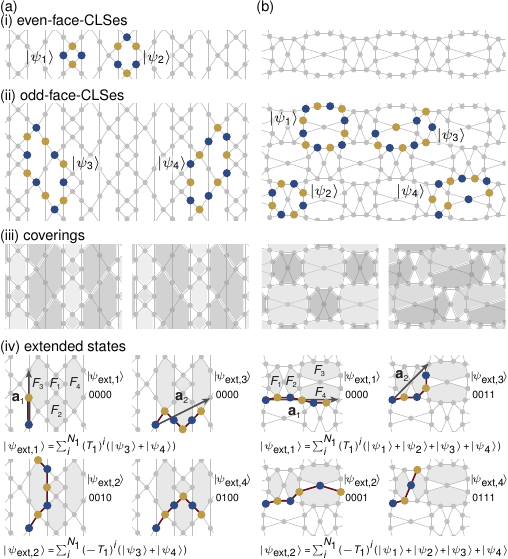}
		\refstepcounter{Sfig}\label{sfig:basis}
		\caption{Example bases for two line-graph lattices: \textbf{(a)} the heptagon-heptagon-hexagon-square kagome lattice and \textbf{(b)} Example 2 from the main text. In both of these cases, the basis is comprised of four CLSes per unit cell, with two linearly dependent CLSes removed from two coverings of the torus by these CLSes, and two linearly independent extended states included. Notice that this yields a total of exactly four CLSes per unit cell, reflecting the fourfold degenerate gapped flat bands. Parts \textbf{(i)} through \textbf{(iv)} depict the steps to construct a basis from the text.}
	\end{figure}
	
	A complete basis for our $1<D\le4$ line-graph lattice flat bands can be defined via the following prescription, with two examples shown in Figure \ref{sfig:basis}. Recall that all of these lattices have root graphs with 0, 1, or 2 even-sided faces per unit cell, and 2 or 4 odd-sided faces per unit cell, seen in Table \ref{stab:geoms}.
	\begin{enumerate}
		\item For each even-sided face, add its corresponding cycle CLS to the basis.
	\end{enumerate}
	In Figure \ref{sfig:basis}(a)(i), the root-graph unit cell has one square face and one hexagon face per unit cell; each of these gives rise to a cycle CLS ($|\psi_1\rangle$ and $|\psi_2\rangle$) in the line-graph lattice.
	In Figure \ref{sfig:basis}(b)(i), there are no even-sided faces.
	\begin{enumerate}
		\item[2a.] If there are two odd-sided faces (per unit cell) then for each odd-sided face and its $C_2$ image, add two corresponding CLSes to the basis. If their boundaries share at least two distinct vertices, two cycle CLSes can be added; if a single vertex, one cycle CLS and one chain CLS can be added; if none, two linearly independent chain CLSes can be added.
	\end{enumerate}
	In Figure \ref{sfig:basis}(a)(ii), the root-graph unit cell has two heptagon faces. Their boundaries share two distinct vertices, hence there are two distinct cycle CLSes (per unit cell) around the two heptagons,  $|\psi_3\rangle$ and $|\psi_4\rangle$.
	\begin{enumerate}
		\item[2b.] If instead there are four odd-sided faces (per unit cell), then for each odd-sided face and its $C_2$ image, add one corresponding CLS to the basis. If the boundary of both faces share at least one vertex, a cycle CLS can be added; otherwise, a chain CLS will be added. Then, for two odd-sided faces of differing number of edges whose boundaries share at least one vertex, add a cycle CLS and its $C_2$ image.
	\end{enumerate}
	In Figure \ref{sfig:basis}(b)(ii), the root-graph unit cell has two heptagon and two pentagon faces. The heptagons share three vertices at their boundaries, and the pentagons share one; hence, we add two cycle CLSes per unit cell to the basis: one enclosing two heptagons  ($|\psi_1\rangle$), and one enclosing two pentagons ($|\psi_2\rangle$). We also add two cycle CLSes per unit cell, which are $C_2$ images of one another and each enclosing one pentagon and one heptagon ($|\psi_3\rangle$ and $|\psi_4\rangle$).
	\begin{enumerate}
		\item[3.] Remove the linearly dependent CLSes from this set. Indeed, after the first two steps, the basis consists of one eigenfunction per face of the root-graph unit cell, for a total of $D$ line-graph eigenfunctions per unit cell, meant to fill the $D$-fold-degenerate flat bands.
	    However, they may not all be linearly independent; as stated in flat-band eigenstate property \ref{itm:FBannihil}, if we have a set of CLSes that completely covers the torus, they will annihilate upon doing so, leaving one linearly dependent state.
	\end{enumerate}
	For example, as seen in Figure \ref{sfig:basis}(a)(iii) there are two ways to cover the torus completely with the cycle CLSes, thus rendering two of these cycle CLSes as linearly dependent and reducing our set of states by two.
	Similarly, in Figure \ref{sfig:basis}(b)(iii) the four cycle CLSes per unit cell can be paired to cover the torus completely in two ways, hence two of the cycle CLSes are linearly dependent and must be removed from our set of states.
	\begin{enumerate}
	    \item[4.] Finally, we find and add the extended states that are linearly independent with the remaining CLSes. We elaborate on this process below.
	\end{enumerate}
	We find linearly independent extended states by first generating linearly dependent extended states via the methods \ref{itm:FBCLSgen} and \ref{itm:FBextgen}.
	For example, in Figure \ref{sfig:basis}(a)(iv), $|\psi_3\rangle$ and $|\psi_4\rangle$ can be used to generate two extended states: $|\psi_{\mathrm{ext},1}\rangle = \sum_i^{N_1} (T_1)^i (|\psi_3\rangle + |\psi_4\rangle)$ (upper left) and $|\psi_{\mathrm{ext},2}\rangle = \sum_i^{N_1} (-T_1)^i (|\psi_3\rangle + |\psi_4\rangle)$ (lower left), where $T_1$ denotes translation by one unit cell in the lattice vector $\mathbf{a}_1$ direction, and there are $N_1$ unit cells in the $\mathbf{a}_1$ direction.
	Similarly, all four cycle CLSes (per unit cell) of Figure \ref{sfig:basis}(b)(iv) generate two extended states: $|\psi_{\mathrm{ext},1}\rangle = \sum_i^{N_1} (T_1)^i (|\psi_1\rangle + |\psi_2\rangle + |\psi_3\rangle + |\psi_4\rangle)$ (Figure \ref{sfig:basis}(b)(iv), upper left) and $|\psi_{\mathrm{ext},2}\rangle = \sum_i^{N_1} (-T_1)^i (|\psi_1\rangle + |\psi_2\rangle + |\psi_3\rangle + |\psi_4\rangle)$ (lower left).
	
	Furthermore, if we now consider the binary string representations of these extended states, adding or subtracting cycle CLSes that come from even-sided faces will result in (linearly dependent) extended states with binary string representations whose leading bits differ.
	Similarly, adding or subtracting chain or cycle CLSes that come from two odd-sided faces will result in (linearly dependent) extended states with binary string representations trailing bits differ in an even number of places.
	From this reasoning, we see that in both of our examples, all extended states in the $\mathbf{a}_1$ lattice direction are linearly dependent with our set of CLSes.
	
	However, in both of our examples, the CLSes cannot be combined to create extended states in the $\mathbf{a}_2$ lattice direction; therefore, we have two extended states in this direction which are linearly independent with our set of CLSes.
	Two possible such states are shown in Figure \ref{sfig:basis}(a)(iv) and (b)(iv), upper and lower right, as $|\psi_{\mathrm{ext},3}\rangle$ and $|\psi_{\mathrm{ext},4}\rangle$.
	From considering the associated bitstrings, again we see that any additional extended states along this lattice direction can be generated through adding or subtracting CLSes to or from  $|\psi_{\mathrm{ext},3}\rangle$ and $|\psi_{\mathrm{ext},4}\rangle$.
	
	\section{From root-graph lattice, to line-graph lattice flat-band representation: proofs}\label{appx:roottolinebr}
	
	In this appendix we prove the key relationships presented in the main text to determine a line-graph lattice's flat-band representation based on its root graph.
	We begin with several properties pertaining to the geometry of our root-graph lattices.
	
	\begin{figure}[tb]
		\centering
		\includegraphics[width=\columnwidth]{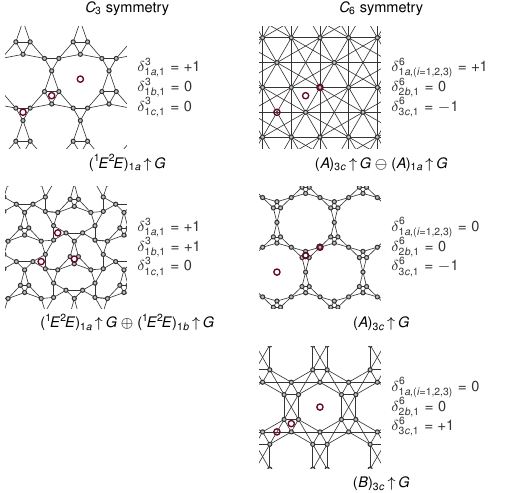}
		\refstepcounter{Sfig}\label{sfig:C3C6}
		\caption{Examples of line-graph lattices with $C_3$ or $C_6$ symmetry and $1 < D \leq 4$. For each lattice, the maximal Wyckoff positions, their associated real-space invariants, and flat-band representations are identified. Notice that faces created from faces in the root graph sit on a maximal Wyckoff position whenever the number of sides shares a common factor (other than 1) with $s$, for point-group symmetry $C_s$ of the lattice.}
	\end{figure}
	
	\begin{figure}[tb]
		\centering
		\includegraphics[width=\columnwidth]{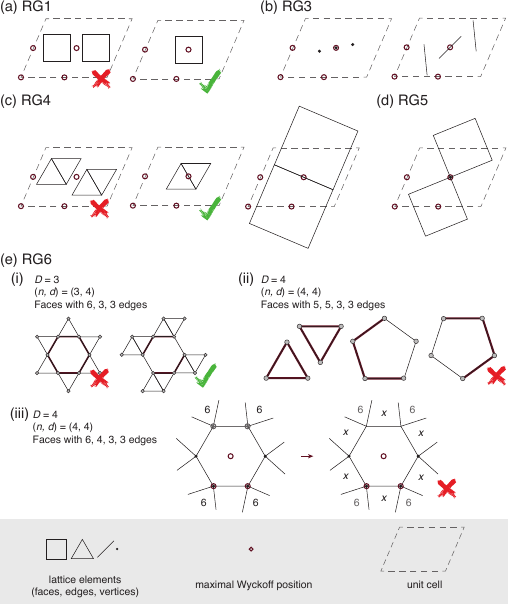}
		\refstepcounter{Sfig}\label{sfig:RGedges}
		\caption{Sketches to assist the arguments presented for properties pertaining to the geometry of our root-graph lattices. \textbf{(a)} Sketch for \ref{itm:RGface}, showing that in our root-graph lattices, faces with an even number of sides must sit on maximal Wyckoff positions. \textbf{(b)} Sketch for \ref{itm:RGparity} showing how vertices and edges on nonmaximal Wyckoff positions must come in pairs, but vertices on and edges on maximal Wyckoff positions do not. \textbf{(c)} Sketch for the argument of \ref{itm:RGsharededgeC2}, showing that two faces that are $C_2$ images of each other and share an edge, must have that edge on a maximal Wyckoff position. \textbf{(d)} Sketch for \ref{itm:RGsharedvertexC2} showing that for two even-sided faces that are $C_2$ images of each other and share a vertex but not an edge, the shared vertex lies on a maximal Wyckoff position. \textbf{(e)(i), (ii), (iii)} Sketch for lattice solutions (1), (2), and (3), respectively, of the arguments for \ref{itm:RGedges}. In (e)(i) and (ii), the red edges highlight edges shared between faces of differing number of sides. In (e)(iii), the two vertices and maximal Wyckoff positions in grey denote those in a neighboring unit cell, ``6''s indicate regions occupied by a hexagonal face, and the ``$x$''s indicate regions occupied by faces of the same number of edges.}
	\end{figure}
	
	\begin{enumerate}
		\item[\namedlabel{itm:RGface}{RG1}] \emph{For a $D\le4$ root-graph lattice with point-group symmetry $C_s$, all faces with a number of sides that shares a common factor (other than $1$) with $s$ will sit at a maximal Wyckoff position.}
	\end{enumerate}
	For $C_3$ and $C_6$, we see this is the case for all such lattices found, shown in Figure \ref{sfig:C3C6}.
	
	For $C_2$, we begin by noting that a face has a $C_2$ center at its center only if it has an even number of sides, because odd-sided faces cannot be invariant under inversion.
	As seen in Table \ref{stab:geoms}, there are no non-bipartite $D=2$ root-graph lattices with even-sided faces, our $D=3$ root-graph lattices must have one even-sided face per unit cell, and our $D=4$ root-graph lattices can have zero or two even-sided faces per unit cell.
	In the $D=4$ lattices with two even-sided faces, these faces do not have the same number of sides.
	
	Now assume for the sake of contradiction that there exists an even-sided face that does not have a $C_2$ center about its center.
	As shown in the left of Figure \ref{sfig:RGedges}(a), take its $C_2$ image about a $C_2$ center; this image is a separate face within the unit cell, creating a total of two even-sided faces with the same number of sides.
	However, we have established that $D=2$ root-graph lattices do not have even-sided faces, $D=3$ root-graph lattices cannot have more than one even-sided face per unit cell, and $D=4$ root-graph lattices cannot have two even-sided faces with an equal number of sides (Table \ref{stab:geoms}).
	Contradiction; for the $C_2$-symmetric lattices considered in this work, all even-sided faces must have a $C_2$ center at its center, and therefore sit at a maximal Wyckoff position (right of Figure \ref{sfig:RGedges}(a)).
	
	\begin{enumerate}
		\item[\namedlabel{itm:RGvertex}{RG2}] \emph{For a vertex to sit on a maximal Wyckoff position with point-group symmetry $C_s$, it must have coordination number a multiple of $s$.}
	\end{enumerate}
	
	This follows directly from the definition of a point-group symmetry $C_s$; under rotation of $2 \pi k/s$ radians about a vertex, $k \in {1, 2, \dots s-1}$, each edge adjacent to that vertex must map onto another edge adjacent to that vertex.
	
	\begin{enumerate}
		\item[\namedlabel{itm:RGparity}{RG3}] \emph{For root-graph lattices with $C_2$ point-group symmetry only, the parity of number of vertices (edges) on maximal Wyckoff positions is equal to the parity of number of vertices (edges) per unit cell.}
	\end{enumerate}
	
	Notice that all vertices (edges) on \emph{non}maximal Wyckoff positions must always come in pairs to preserve $C_2$ symmetry, see Figure \ref{sfig:RGedges}(b).
	The number of vertices (edges) per unit cell is given by the number of vertices (edges) on maximal Wyckoff positions, plus the number of vertices (edges) on non-maximal Wyckoff positions, hence the claim must be true.
	As seen in Table \ref{stab:geoms}, for $D=4$ (and for even $D$ more generally) the number of vertices (edges) per unit cell is always even; thus the number of vertices (edges) on maximal Wyckoff positions is also always even.
	
	\begin{enumerate}
		\item[\namedlabel{itm:RGsharededgeC2}{RG4}] \emph{For $D\le4$ root-graph lattices with $C_2$ point-group symmetry only, if two faces that are $C_2$ images of each other share a single edge, then there is a $C_2$ center located at the middle of the shared edge.}
	\end{enumerate}
	
	This lemma will be useful in proving \ref{itm:RGedges}.
	
	Consider two faces that are $C_2$ images of each other and share a single edge.
	First consider the case where these faces are odd-sided.
	Then they cannot map onto each other under translation because odd-sided faces are not invariant under $C_2$.
	Thus they can be defined to be in the same unit cell.
	Now assume for the sake of contradiction that their shared edge does not cross a maximal Wyckoff position (left of Figure \ref{sfig:RGedges}(c)).
	In this case, rotate the pair about a maximal Wyckoff position---this will result in four copies of the face per unit cell.
	As seen in Table \ref{stab:geoms}, there are no such $D\le4$ root-graph lattices with four faces of the same number of edges in the unit cell; we have a contradiction, and their shared edge must go through a maximal Wyckoff position.
	
	Now consider the case where the two faces are instead even-sided.
	They must each be centered about a maximal Wyckoff position according to property \ref{itm:RGface} of our root-graph lattices (see right of Figure \ref{sfig:RGedges}(c)).
	From Table \ref{stab:geoms}, we also know they cannot sit in the same unit cell because they are even-sided and both have the same number of sides; hence they sit in adjacent unit cells.
	Then the midpoint of their centers is also a maximal Wyckoff position, which is necessarily on the shared edge.
	
	\begin{enumerate}
		\item[\namedlabel{itm:RGsharedvertexC2}{RG5}] \emph{In a $C_2$-symmetric $D\le4$ root-graph lattice, if two even-sided faces that are $C_2$ images of each other share a vertex (but not an edge), then the vertex is on a $C_2$ center.}
	\end{enumerate}

	This is a corollary to \ref{itm:RGsharededgeC2} and will also be useful in proving \ref{itm:RGedges}.
	
	The even-sided faces must each be centered about a maximal Wyckoff position (property \ref{itm:RGface}), and cannot sit in the same unit cell (Table \ref{stab:geoms}).
	Thus they must sit in adjacent unit cells, and the midpoint of their centers is also a maximal Wyckoff position, which is on the shared vertex, see Figure \ref{sfig:RGedges}(d).
	
	\begin{enumerate}
		\item[\namedlabel{itm:RGedges}{RG6}] \emph{For $D\le4$ root-graph lattices with $C_2$ point-group symmetry only, there are always edges on at least two maximal Wyckoff positions.}
	\end{enumerate}

	First consider root-graph lattices with odd coordination number.
	Then from \ref{itm:RGvertex}, we know that none of the vertices can sit on a maximal Wyckoff position.
	Similarly, from \ref{itm:RGface} we know that only even-sided faces will be centered on a maximal Wyckoff position.
	Then, because there are a total of four maximal Wyckoff positions for $C_2$-symmetric lattices, and these $D\leq 4$ root-graph lattices can have at most two even-sided faces per unit cell (Table \ref{stab:geoms}), there must be edges sitting on at least two maximal Wyckoff positions.
	
	If instead we have a root-graph lattice with even coordination number, from Table \ref{stab:geoms} we find only a handful of possible solutions: (1) $D=3$, $(n,d)=(3,4)$, with faces of 6, 3, and 3 edges; (2) $D=4$, $(n, d)=(4, 4)$, with faces of 5, 5, 3, and 3 edges; and (3) $D=4$, $(n, d)=(4, 4)$, with faces of 6, 4, 3, and 3 edges.
	
	First, we will show that lattice solution (2) must have edges sitting on at least two maximal Wyckoff positions.
	If each edge of the two triangle faces is shared with an edge of a pentagon face, notice in Figure \ref{sfig:RGedges}(e)(ii) that there will be a total of four unpaired edges among the pentagons.
	Thus, they must share at least two edges; from \ref{itm:RGsharededgeC2}, we conclude that lattices from solution (2) have edges on at least two maximal Wyckoff positions.
	
	Second, we consider lattice solution (1) and proceed in a similar manner.
	If each edge of the two triangle faces is shared with an edge of the hexagon face, as shown in Figure \ref{sfig:RGedges}(e)(i) we create the kagome lattice, exhibiting $C_6$ symmetry.
	If another lattice geometry exists for this solution, the two triangle faces must then share at least one edge.
	Then if they share one edge, then there will be two unpaired edges for the hexagon, and the hexagon must also share an edge with its translated copy in a neighboring unit cell (see Figure \ref{sfig:RGedges}(e)(i)); notice that this lattice has $C_2$ point-group symmetry only, and not $C_6$.
	Hence, the root-graph lattice from solution (1) with only $C_2$ symmetry also has edges on at least two of its maximal Wyckoff positions.
	
	Third, for lattice solution (3) we consider the hexagon face and try to place the square face and two triangle faces around it to construct the lattice unit cell.
	We assume for the sake of contradiction that there exists a construction where there are edges on fewer than two maximal Wyckoff positions.
	In this lattice solution, there are only $n=4$ vertices per unit cell; as a result, two pairs of vertices at the boundary of the hexagon must be copies of each other, translated by a lattice vector.
	In other words, these vertices must each be at the boundary of two copies of the hexagon face.
	At the same time, the hexagon face cannot share an edge with its translated copy.
	If it does, then this edge will lay on a maximal Wyckoff position (by \ref{itm:RGsharededgeC2}); then by \ref{itm:RGparity}, there must be at least two edges on maximal Wyckoff positions, violating the conditions of our assumption.
	Therefore, the two vertices that are adjacent to two copies of the hexagon face are on maximal Wyckoff positions, as shown in \ref{sfig:RGedges}(e)(iii).
	The other two faces that these vertices are adjacent to must then be $C_2$ images of each other and have the same number of sides.
	However, this results in six copies of the face, all sharing distinct edges with the hexagon.
	The square only has four sides, so they cannot be squares; if they are triangles, then all triangles only share edges with hexagons, and the squares cannot be included in the unit cell.
	Thus we find the assumption must be false; there must be edges on two of the maximal Wyckoff positions.
	
	\begin{enumerate}
		\item[\namedlabel{itm:RGC3C6}{RG7}] \emph{The $C_3$- and $C_6$-symmetric root-graph lattices have maximal Wyckoff positions as tabulated in Table 1, and RSIs as stated in the main text.}
	\end{enumerate}
	We show them directly in Figure \ref{sfig:C3C6}.\newline
	
	In this Appendix thus far, we have established several relationships for root-graph lattices between their geometric properties and graph elements at high-symmetry points.
	From Appendix \ref{appx:rootlineprop}, we know how these root-graph graph elements relate to the graph elements of the corresponding line graphs (properties \ref{itm:LGsubgraph} and \ref{itm:LGcycle}).
	We now show the relationship between the line-graph graph elements at high-symmetry points of the line-graph lattice and the RSIs of those points, for lattices with $C_2$ symmetry.
	
	\begin{figure}[tb]
		\centering
		\includegraphics[width=\columnwidth]{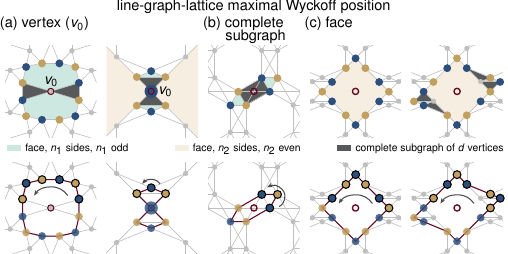}
		\refstepcounter{Sfig}\label{sfig:C2RSIs}
		\caption{Determination of real-space invariants (RSIs) for $C_2$-symmetric line-graph lattices as stated in \ref{itm:RSI}, determined by constructing $C_2$ eigenstates from flat-band energy eigenstates and computing the number of $(+1)$-eigenvalue eigenstates relative to the number of $(-1)$-eigenvalue eigenstates. Blue (yellow) circles denote positive (negative) relative wavefunction amplitude, and larger circles have twice the amplitude of the smaller circles. Even cycles corresponding to the CLSes are highlighted in red. We find that the RSI values depend on whether the maximal Wyckoff position is \textbf{(a)} occupied by a vertex, \textbf{(b)} at the center of a complete subgraph, or \textbf{(c)} at the center of a face of the corresponding line-graph lattice. Specific line-graph lattices and eigenfunctions are shown here for concreteness, although the claims are general.}
	\end{figure}
	
	\begin{enumerate}
		\item[\namedlabel{itm:RSI}{RSI1}] \emph{For a line-graph lattice with $C_2$ point-group symmetry only, the RSI $\delta^2_{w, 1} \equiv m^2_{w, 1} - m^2_{w, 0}$ is equal to $-1$ for maximal Wyckoff positions $w$ occupied by a vertex, $+1$ for those at the center of complete subgraph, and $0$ for those at the center of a face.}
	\end{enumerate}
	Given a maximal Wyckoff position, we need only look at $C_2$ eigenfunctions locally around that position in the line-graph lattice.
	These positions can be occupied by vertices, complete subgraphs, or faces, because each of these graph elements results directly from an edge, vertex, or face of the root-graph lattice per the line graph construction (as stated in properties \ref{itm:LGsubgraph} and \ref{itm:LGcycle}).
	
	First, consider maximal Wyckoff positions of line-graph lattices that are occupied by a vertex $v_0$, see Figure \ref{sfig:C2RSIs}(a).
	Two faces created from faces in the root graph, which share the edge in the root graph that $v_0$ originated from, will share $v_0$.
	This vertex $v_0$ will also be shared by two complete subgraphs (colored in dark grey), which are created from the vertices in the root graph that are at either end of the edge that $v_0$ originated from.
	If the faces are odd-sided (colored in green, see left of Figure \ref{sfig:C2RSIs}(a)), we can define a flat-band $C_2$ eigenstate $|\phi\rangle$ as a cycle CLS from the even cycle around the two faces (Appendix \ref{appx:eigenfn}).
	This eigenstate has $C_2$ eigenvalue $+1$ with respect to the maximal Wyckoff position under consideration; notice that this is because the even cycle (highlighted in red) will always have vertices given by all vertices at the boundary of the odd-sided faces, minus the shared vertex on the maximal Wyckoff position, as drawn in the bottom left of Figure \ref{sfig:C2RSIs}(a).
	Then there must be an even number of vertices where $|\phi\rangle$ has nonzero amplitude at the boundary of each odd-sided face; because the amplitudes are real-valued and alternate in sign, $|\phi\rangle$ will then have $+1$ as its $C_2$ eigenvalue.
	
	If the two faces sharing $v_0$ are instead even-sided (see right of Figure \ref{sfig:C2RSIs}(a), colored in orange), then each of the even-sided faces can be used to define a cycle CLS $|\phi'\rangle$.
	Because these are not $C_2$ eigenstates individually, under the construction given by Equation \ref{eq:CsEigenfn} of the main text they create $C_2$ eigenstates $|\phi'_{k=0}\rangle$ and $|\phi'_{k=1}\rangle$ of eigenvalues $+1$ and $-1$, respectively, and do not contribute to the RSI.
	Consequently we consider two other faces that are odd-sided and $C_2$ images of one another (colored in green); these must exist as seen in Table \ref{stab:geoms}.
	Then we can define a flat-band $C_2$ eigenstate $|\phi\rangle$ as a chain CLS from the two odd-sided faces, as in the right of Figure \ref{sfig:C2RSIs}(a).
	Here, too, $|\phi\rangle$ has $C_2$ eigenvalue $+1$ with respect to the maximal Wyckoff position $C_2$ center, resulting from the fact that the eigenfunction must have non-zero amplitude on the vertex located on the maximal Wyckoff position (see bottom right, Figure \ref{sfig:C2RSIs}(a)).
	
	For other $C_2$ eigenstates at this maximal Wyckoff position, we find an equal number of them with eigenvalue $+1$ and eigenvalue $-1$.
	In particular, CLSes originating from even-sided faces or pairs of odd-sided faces not centered about the maximal Wyckoff position must be combined with their $C_2$ image to make a $C_2$ eigenfunction.
	As a result, these eigenfunctions do not contribute to the RSI.
	Any remaining $C_2$ eigenstates can be written as linear combinations of these eigenstates and $|\phi\rangle$.
	
	With a similar argument, we next consider the maximal Wyckoff positions occupied by a complete subgraph (colored in dark grey), see Figure \ref{sfig:C2RSIs}(b).
	In the line graph, there must be (at least) two odd-sided faces (colored in green) created from faces in the root-graph lattice that are $C_2$ images of one another and share vertices with the complete subgraph.
	Then we can define a flat-band $C_2$ eigenstate $|\phi\rangle$ as in Figure \ref{sfig:C2RSIs}(b).
	This eigenstate will have $C_2$ eigenvalue $-1$, because as shown in the bottom half of Figure \ref{sfig:C2RSIs}(b), now the vertices of the even cycle (highlighted in red) will be given by all vertices at the boundary of both odd sided faces.
	Hence, there are an odd number of vertices at the boundary of each odd-sided face where $|\phi\rangle$ has nonzero amplitude; because the amplitudes are real-valued and alternate in sign, $|\phi\rangle$ will then have $-1$ as its $C_2$ eigenvalue.
	As before, all other $+1$- and $-1$-eigenvalued $C_2$ eigenfunctions will be equal in number and will not contribute to the RSI value.
	
	Third, we consider the maximal Wyckoff positions occupied by a face (created from a face in the root graph).
	This face must be even-sided (\ref{itm:RGface}), and can be used to define a cycle CLS $|\phi^\alpha\rangle$, as shown in Figure \ref{sfig:C2RSIs}(c), left.
	In this example, $|\phi^\alpha\rangle$ has $C_2$ eigenvalue $+1$; more generally the eigenvalue will be equal to $+1$ if the number of vertices at the boundary of the face is divisible by $4$, otherwise $-1$.
	This can be seen by considering half of the boundary (Figure \ref{sfig:C2RSIs}(c), bottom left); if there are an even number of vertices in this half (\emph{i.e.} the total number of vertices is divisible by 4), then as the wavefunction alternates in amplitude on these vertices, it will have the same amplitude upon reaching the second half of the boundary.
	Otherwise, it will have opposite amplitude, for $C_2$ eigenvalue $-1$.
	
	In this case, there exists a second flat-band $C_2$ eigenfunction that also contributes to the RSI.
	More specifically, there must be (at least) two odd-sided faces that are $C_2$ images of each other (Table \ref{stab:geoms}) and share vertices with the even-sided face sitting on the maximal Wyckoff position; a cycle CLS $|\phi^\beta\rangle$ can be constructed from the even cycle encircling these three faces.
	As seen in Figure \ref{sfig:C2RSIs}(c), right, $|\phi^\beta\rangle$ has $C_2$ eigenvalue $-1$.
	More generally, the vertices of the even cycle where $|\phi^\beta\rangle$ has nonzero amplitude are given by all vertices at the boundary of the even-sided face and both odd-sided faces, minus the two shared vertices between each of the odd-sided faces and the even-sided face.
	As a result, when we again consider half of the boundary, as in the bottom right of Figure \ref{sfig:C2RSIs}(c), we find an additional number of vertices given by two less than the number of vertices at the boundary of the odd-sided face.
	This additional number is thus always odd; therefore, the $C_2$ eigenvalue of $|\phi^\beta\rangle$ will always be opposite that of $|\phi^\alpha\rangle$.
	Here, too, other $C_2$ eigenstates will not contribute to the RSI.
	
	As a result, if the line-graph lattice has a vertex on a maximal Wyckoff position, there will be one more $(+1)$-eigenvalued flat-band $C_2$ eigenfunction than $(-1)$-eigenvalued, and the RSI $\delta^2_{w, 1} = -1$.
	If the lattice instead has a complete subgraph on the maximal Wyckoff position, there will be one more $(-1)$-eigenvalued flat-band $C_2$ eigenfunction than $(+1)$-eigenvalued, and $\delta^2_{w, 1} = +1$.
	If the lattice instead has a face on the maximal Wyckoff position, there will be an equal number of $(+1)$- and $(-1)$-eigenvalued flat-band $C_2$ eigenfunctions, hence $\delta^2_{w, 1} = 0$.
	
	Once the RSI values have been determined, representations can be determined via Equation \ref{eq:summ} of the main text.
	For our $D=2$ line-graph lattices with gapped flat bands, we only find two lattices: Example 1 of the main text (with $C_6$ symmetry), and the line graph of the nonagon-triangle lattice (with $C_3$ symmetry), which is presented and analyzed in full in Appendix \ref{appx:exs}.
	In the first column of Tables \ref{tab:D=3splitBR} and \ref{tab:D=4splitBR}, we tabulate the possible flat-band representations for our $D=3$ and $D=4$ line-graph lattices that exhibit $C_2$ symmetry.
	Notice that all of these possible representations are sums of EBRs.
	
	\section{Perturbations to split the degeneracy of $D\ge3$ line-graph-lattice flat bands}\label{appx:perturb}
	
	Here we prescribe how to split the degeneracy of the $(D=3)$- and $(D=4)$-fold degenerate flat bands at energy $-2$ to result in two-fold-degenerate gapped flat bands at $-2$.
	We consider perturbations \cite{Leykam2018s} consisting of on-site energies or additional hoppings that preserve the original rotational and translational symmetries of the line-graph lattice.
	
	By considering only symmetry-preserving perturbations, we place constraints on the minimum number of on-site energies or additional hoppings required.
	More specifically, a single on-site energy (per unit cell) will be symmetry-preserving only if the vertex sits on a maximal Wyckoff position with the same point-group symmetry as the lattice.
	While we find $C_2$-symmetric line-graph lattices where such vertices exist, \emph{e.g.} Example 2, these vertices do not exist in the $C_3$ and $C_6$ lattices we consider, see Figure \ref{sfig:C3C6}.
	In these lattices, we require a minimum of three or six on-site energies to preserve the lattice symmetry.
	
	Similarly, because edges are invariant under $C_2$ about their centers, to preserve the lattice symmetry $C_s$, new hopping perturbations must be added in groups of size equal to half of the least common multiple of $2$ and $s$.
	More explicitly, a single new hopping (per unit cell) can only be added to lattices with $C_2$ symmetry; the same is true for pairs of new hoppings.
	Furthermore, single new hoppings must pass through the center of a face that sits on a maximal Wyckoff position, and hopping pairs must be $C_2$ images of one another.
	For lattices with $C_3$ and $C_6$ symmetry, a minimum number of three new hoppings is required.
	
	For any given lattice and perturbation, the perturbation's effect on the flat bands can of course be computed directly through examining the band structure or computing the rank of the momentum-space matrix $H(k) + 2I + H'(k)$ for momentum $k$, where $H(k)$ is the matrix Hamiltonian, $I$ is the identity matrix of the same size (and shifts the flat bands to zero energy), and $H'(k)$ is the perturbation matrix.
	In the matrix rank approach, the flat-band band degeneracy is equal to the nullity (dimension minus rank) of $H(k) + 2I + H'(k)$.
	In general, the more non-zero elements there are in $H'(k)$, the smaller the nullity of $H(k) + 2I + H'(k)$.
	This suggests that in general, the addition of too many perturbations leads to fewer than two flat bands at $-2$.
	Indeed, for our line-graph lattices exhibiting $C_3$ and $C_6$ symmetry, we do not find any symmetry-preserving perturbations which split the bands successfully and note that an increased number of perturbations is required to maintain these symmetries.
	We therefore discuss lattices with these symmetries separately at the end of this appendix.
	
	Thus we focus on line-graph lattices with only $C_2$ symmetry.
	We determine the consequences of various perturbations on the real-space flat-band eigenstates: cycle CLSes, chain CLSes, and extended states.
	Because the number of such linearly independent states is directly related to the number of flat bands \cite{Bergman2008, Kollar2019}, we can then draw connections between features of the perturbations and resulting band splitting.
	
	In particular, recall from Appendix \ref{appx:eigenfn} that the flat-band eigenstates require compact support: sites where the wavefunction is zero-valued will remain zero-valued upon applying the Hamiltonian, due to complete destructive interference of wavefunction amplitudes after hopping.
	Recall also that there is a subset of $D$ real-space flat-band eigenstates per unit cell that are linearly independent and can be used to define a complete basis for the $D$-fold-degenerate flat-band states in the unperturbed line-graph lattice.
	Upon perturbation, we will show that some of these eigenstates will have a modified energy or lose compact support while other wavefunctions will maintain their compact support.
	By counting the maximum number of linearly independent flat-band states possible for the perturbed lattice, we determine the change in the flat-band band degeneracy and whether the remaining flat bands (if they exist) are gapped \cite{Bergman2008, Kollar2019}. 
	In particular, a perturbation that results in two-fold-degenerate gapped flat bands at the flat-band energy $-2$, must result in a lattice that provides compact support to exactly two flat-band eigenstates per unit cell that are all linearly independent.

	\subsection{On-site energy perturbations for $C_2$-symmetric lattices}
	
\begin{figure}[tb]
		\centering
		\includegraphics[width=\columnwidth]{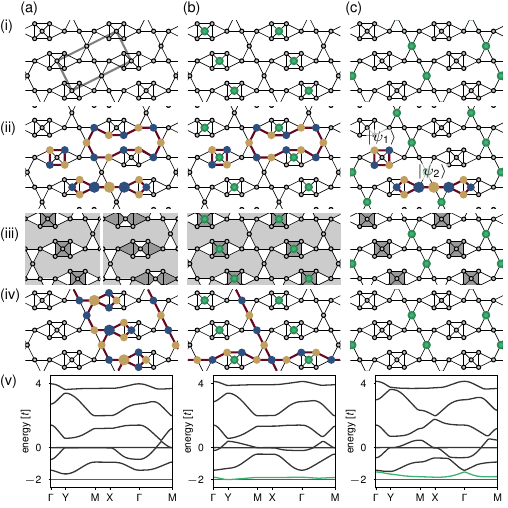}
		\refstepcounter{Sfig}\label{sfig:onvertexpert}
		\caption{The line graph of the dodecagon-triangle-triangle lattice with on-site energy perturbations. \textbf{(a)} Before perturbation. \textbf{(b)} An on-site energy perturbation (green circles) that unsuccessfully splits the bands. \textbf{(c)} An on-site energy perturbation (green circles) that successfully splits the bands. In each subfigure, \textbf{(i)} shows the lattice (in (a), a single unit cell is outlined in grey); \textbf{(ii)} shows cycle and chain CLS energy eigenstates, each highlighted with red lines, that are used in constructing the basis; \textbf{(iii)} shows whether the CLSes cover the torus; and \textbf{(iv)} shows extended states that are linearly independent with the set of CLSes (plus translated copies) shown in (ii). Blue (yellow) circles denote positive (negative) relative wavefunction amplitude, and larger circles have twice the amplitude of the smaller circles. In \textbf{(v)}, we show the band spectrum. Flat bands are in blue, and flat bands that have been perturbed to become dispersive are in green.}
	\end{figure}

	We have noted in the main text that on-site energy perturbations cannot change the topology of the split bands relative to the original topology of the bands prior to perturbation.
	However, it can be instructive to see how on-site energy perturbations can be used to split the band degeneracy.
	An example is depicted in Figure \ref{sfig:onvertexpert}, where $D=3$.
	
	In this example, pre-perturbation we can construct the flat-band basis via the procedure described in Appendix \ref{appx:eigenfn} and Figure \ref{sfig:basis}.
	This construction begins with two cycle CLSes and one chain CLS per unit cell (Figure \ref{sfig:onvertexpert}(a)(ii)).
	This results in two complete coverings of the torus ((a)(iii)), removing two CLSes from our set (from property \ref{itm:FBannihil}); at the same time, we have two extended states (Figure \ref{sfig:onvertexpert}(a)(iv)) that are linearly independent from our set.
	In total, we find 3 real-space flat-band eigenstates per unit cell that are linearly independent, consistent with having 3-fold-degenerate gapped flat bands as shown in blue in Figure \ref{sfig:onvertexpert}(a)(v).
	
	Now in Figure \ref{sfig:onvertexpert}(b) we introduce an on-site energy.
	This modifies the energy of any CLSes---namely, the chain CLS of a(ii)---that have nonzero amplitude on the vertex with on-site energy.
	However, the two cycle CLSes per unit cell of (a)(ii) remain unaffected, and can be used as a starting point in constructing the flat-band basis.
	Because they cover the torus ((b)(iii), \ref{itm:FBannihil}), one of these cycle CLSes is linearly dependent with the others and must be removed from the set as before, for a running total of $2M-1$ flat-band states given a lattice with $M$ unit cells.
	We additionally find two extended states that will be linearly independent with the set, for a total of $2M+1$ flat-band states.
	Because $2M$ states fit in two flat bands, our total number of flat-band states indicates that the perturbed spectrum contains two flat bands (at energy $-2$) with a band touch to a third, dispersive band.
	Indeed, in Figure \ref{sfig:onvertexpert}(b) we see two flat bands at energy $-2$, touching a dispersive band at the $Y$ point.

	By contrast, in Figure \ref{sfig:onvertexpert}(c) the on-site-energy perturbation yields gapped doubly degenerate flat bands ((c)(v)).
	Here, the cycle CLS (in each unit cell) that encloses the dodecagon will have nonzero amplitude on the vertices with on-site energy, so its energy will be shifted away from $-2$.
	The set of linearly independent flat-band states then consists of one cycle CLS per unit cell $|\psi_1\rangle$ and one chain CLS $|\psi_2\rangle$ per unit cell, as drawn in Figure \ref{sfig:onvertexpert}(c)(ii).
	Because the dodecagon face is not enclosed by any of the CLSes in our set, there is no complete covering of the torus ((c)(iv)).
	There are also no linearly independent extended states; those that encircle the torus in the horizontal ($\mathbf{a}_1$) direction are given by $\sum_i^{N_1} (T_1)^i (\pm|\psi_1\rangle-|\psi_2\rangle)$, while none encircle the torus in the vertical direction and have zero amplitude on the vertices with on-site energy.
	With a total of $2M$ flat-band states, we find the perturbed band spectrum contains two flat bands, gapped from the rest of the spectrum.
	
	Among all $D=3$, $C_2$-symmetric line-graph lattices, the one shown in Figure \ref{sfig:onvertexpert} is the only one we found containing vertices that are adjacent to two copies of the even-sided face inherited from the root graph.
	We conjecture there are no others.
	For all other $D=3$ lattices with $C_2$ symmetry, we conjecture that there is no symmetry-preserving on-site perturbation that can split the bands into two-fold-degenerate gapped flat bands and a dispersive band, because the flat-band basis construction will proceed similarly to that of Figure \ref{sfig:onvertexpert}(b).
	We additionally conjecture that for lattices with $D=4$ (and $C_2$ symmetry), one must apply on-site energies to two or more vertices, and that such a perturbation always exists.
	More specifically, for 4o lattices, the perturbation can be placed on two vertices that sit on maximal Wyckoff positions and are adjacent to the same odd-sided face; for 2e2o lattices, the perturbation can be placed on the two vertices adjacent to both even-sided faces.
	Because these split bands will admit Wannier representations, we find it beyond the motivation of this work to prove these results here.

	\subsection{Hopping perturbations for $C_2$-symmetric lattices}
	
	We begin by noting that we cannot introduce hopping between one vertex in a unit cell and its translated counterparts in neighboring unit cells.
	If we do, then the perturbation Hamiltonian $H'(k)$ will consist of a $\cos(k)$ term on its diagonal, and there will be a quasimomentum $k_0$ for which this perturbation vanishes, yielding a band touching at $k_0$.
	Similarly, if we introduce a hopping between two vertices $v_{L(x), 1}$ and $v_{L(X),2}$ within one unit cell, we cannot also introduce a hopping between $v_{L(X), 1}$ and $T_1 v_{L(X), 2}$, where $T_1$ is translation by one of the lattice vectors.
	Otherwise, the corresponding perturbation Hamiltonian consists of a $\cos(k)$ term on its off-diagonal, and there will be a quasimomentum $k_0$ for which this perturbation vanishes, yielding a band touching at $k_0$.
	
	Even so, we find a class of hopping perturbations that splits the band degeneracy for $D=4$ line-graph lattices with $C_2$ symmetry, and present arguments for why no such perturbations seem to exist for $D=3$ line-graph lattices.
	As a reminder, we consider only hopping perturbations that preserve the lattice symmetry.
	
	\subsubsection{Degeneracy $D=4$}
	
	Now we present hopping perturbations that split the bands for $D=4$.
	These perturbations will be specific to each of the two families (2e2o, where the root-graph lattice unit cell has two even-sided and two odd-sided faces, and 4o, where the root-graph lattice unit cell has four odd-sided faces) but otherwise general across all line-graph lattices within the family.
	
	\begin{figure}[tb]
		\centering
		\includegraphics[width=\columnwidth]{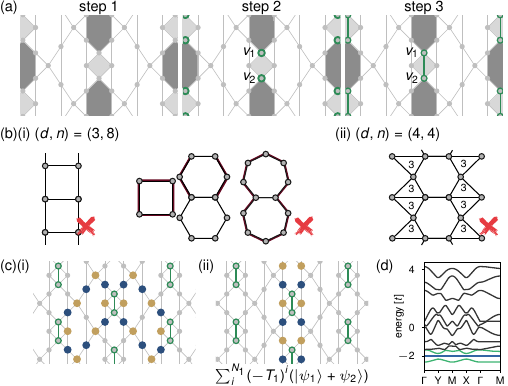}
		\refstepcounter{Sfig}\label{sfig:D=42e2o}
		\caption{Constructing hopping perturbations for $D=4$ line-graph lattices in the 2e2o family, to split the fourfold-degenerate gapped flat band at energy $-2$ and create two-fold-degenerate gapped flat bands at the same energy. \textbf{(a)} The steps (see text) of the 2e2o hopping perturbation construction for the heptagon-heptagon-hexagon-square kagome line-graph lattice, which is in the 2e2o family. Perturbations are in green. \textbf{(b)} Schematic sketches as visual aids to the proof that our constructed 2e2o hopping perturbation always exists. \textbf{(i)} Sketches for root-graph lattices with $(d, n)=(3, 8)$, with the specific example of two heptagon faces, one hexagon face, and one square face per unit cell. Red edges highlight the mismatch in number of edges around the even faces and odd faces in the argument presented in the text. \textbf{(ii)} Sketch for root-graph lattices with  $(d, n)=(4, 4)$, with the specific example of one hexagon face, one square face, and two triangle faces per unit cell. \textbf{(c)(i)} Example flat-band basis after perturbation, comprised of two CLSes per unit cell. Flat-band wavefunctions are shown as blue and yellow circles denoting real-valued relative wavefunction amplitudes of positive and negative sign, respectively. An example eigenstate basis for this lattice before perturbation is in Figure \ref{sfig:basis}. \textbf{(ii)} Linear combination of cycle CLSes from even-sided faces which remains a flat-band eigenstate of the lattice post-perturbation, even though individual cycle CLSes do not. This eigenfunction is linearly dependent with the states (plus translation) in (i). \textbf{(d)} Band spectrum of the perturbed lattice, where blue lines denote flat bands while green lines denote dispersive bands that were flat and at energy $-2$ prior to perturbation.}
	\end{figure}
	
	{\bf 2e2o line-graph lattices:}
	For 2e2o line-graph lattices, we find a hopping perturbation that splits the bands into two gapped flat bands at energy $-2$ and two dispersive bands.
	Its construction is shown schematically in Figure \ref{sfig:D=42e2o}(a) for the line graph of the heptagon-heptagon-hexagon-square lattice and constructed as follows:
	\begin{enumerate}
		\item Identify the two even-sided faces per unit cell, shaded in light and dark gray in Figure \ref{sfig:D=42e2o}(a).
		\item These two faces will have two vertices ($v_1$ and $v_2$) that are at the boundaries of both faces, see the middle image of Figure \ref{sfig:D=42e2o}(a).
		\item Add a hopping between these two vertices, which will cross through one of the even-sided faces. Note that there are two possible such hoppings, \emph{e.g.} between $v_1$ and $v_2$, or between $T_1 v_1$ and $v_2$. In Figure \ref{sfig:D=42e2o}(a), the hopping is drawn as a green edge between $v_1$ and $v_2$.
	\end{enumerate}
	
	These two vertices always exist; by considering the root-graph lattice, this claim is equivalent to the claim that there are two edges that are at both boundaries of the two even-sided faces.
	From Table \ref{stab:geoms}, given that $D=4$, we have $(d,n)=(3,8)$ or $(4, 4)$.
	Because the even-sided faces must sit on maximal Wyckoff positions (\ref{itm:RGface}), there must be an even number of edges that are at both boundaries of the two even-sided faces.
	Then we assume for the sake of contradiction that there are no edges at the boundaries of both even-sided faces, and consider each $(d,n)$ pair.
	
	If the root graph has $(d,n)=(3,8)$, then the smallest even-sided faces possible are of $4$ and $6$ sides.
	If these faces do not share edges, they also cannot share vertices because of the graph is only of degree $d=3$.
	Furthermore, because $d=3$, the square cannot share edges with translated copies of itself, otherwise it cannot be part of the unit cell with the other faces (see Figure \ref{sfig:D=42e2o}(b)(i), left).
	Similarly, the hexagon can share at most one edge with a translated copy of itself, in which case this edge must sit on a maximal Wyckoff position.
	We first assume that this is the case.
	The other two faces of this lattice have $7$ sides each; they must share an edge to occupy the last maximal Wyckoff position (\ref{itm:RGedges}).
	As seen in Figure \ref{sfig:D=42e2o}(b)(i), right, this leaves 13 edges at the boundaries of the two heptagon faces, and a total of 8 edges at the boundary of the square and hexagon faces.
	These numbers of edges are unequal, hence there is no way to match them up so that each of them lies at the boundary of one even-sided face and one odd-sided face; thus, the heptagons must share at least one additional edge.
	This additional edge must sit on a maximal Wyckoff position (\ref{itm:RGsharededgeC2}), yet all four maximal Wyckoff positions are already occupied.
	Therefore, the hexagon cannot share an edge with a translated copy of itself.
	In this case, however, we require a total of $6+4=10$ (unique) vertices per unit cell, but there are only $n=8$ available.
	Therefore, the assumption is false and the square and hexagon must share at least two edges between them (that is, each square shares an edge with a hexagon, as well as an edge with a translated copy of the hexagon).
	
	For the other $(d,n) = (3, 8)$ root graphs, similar arguments can be applied.
	In particular, even if there are shared edges between the boundaries of two copies of the same even-sided face, the number of unique vertices required exceeds $n=8$.
	Thus, for these lattices, there must be shared edges at the boundaries of both even-sided faces.
	
	If instead the root graph has $(d,n)=(4,4)$, from Table \ref{stab:geoms} we see that it must have faces of $3$, $3$, $4$, and $6$ sides.
	From the proof of \ref{itm:RGedges}, we know that the hexagon must share at least one side with a translated copy of itself.
	Then if we place the triangles to avoid having shared edges between the boundaries of the hexagon and square (see Figure \ref{sfig:D=42e2o}(b)(ii), right), we find that the triangles share an edge with each other and the remaining four edges with the hexagon, leaving nowhere for the square in the unit cell.
	Therefore, the assumption must be false and there must be a shared edge between the boundaries of the hexagon and square.
	
	Now we show that our constructed hopping perturbation for 2e2o line-graph lattices splits the band degeneracy and yields two gapped flat bands at energy $-2$ by counting the maximum number of linearly independent flat-band eigenstates.
	In Figure \ref{sfig:D=42e2o}(c)(i) we show the basis construction (one state per unit cell is shown) for our example 2e2o lattice after perturbation; recall that in Figure \ref{sfig:basis}(a) we have constructed a complete basis when there is no perturbation.
	First, notice that the single hopping perturbation removes the compact support on both vertices involved in the perturbation, such that the eigenbasis spanning the post-perturbation subspace of energy $-2$ states cannot have any amplitude on these vertices.
	As a result, the two cycle CLSes for the even-sided faces in the unperturbed lattice (\emph{e.g.} $|\psi_1\rangle$ and $|\psi_2\rangle$ of Figure \ref{sfig:basis}(a)(i)) are not flat-band eigenstates in the perturbed lattice.
	Similarly, the extended states with nonzero amplitude on the perturbed vertices will no longer be flat-band eigenstates of the perturbed lattice (in Figure \ref{sfig:basis}(a)(iv), we can see that these include all extended states along the $\mathbf{a}_2$ direction of the lattice).
	However, the compact support is maintained for the two CLSes from the two odd-sided faces (\emph{e.g.} $|\psi_3\rangle$ and $|\psi_4\rangle$ of Figure \ref{sfig:basis}(a)(ii)), and they remain flat-band eigenstates of the perturbed lattice.
	
	The only linear combinations of the two even-sided-face cycle CLSes that maintain compact support are ones which have zero amplitude on the perturbed vertices.
	Thus they are given by $\sum_i^{N_1} (-T_1)^i (|\psi_1\rangle + |\psi_2\rangle)$, which can be written as pairs of extended states at the (extended) boundaries between the even-sided faces and odd-sided faces (see Figure \ref{sfig:D=42e2o}(c)(ii)).
	Each of the extended states in this pair are linearly dependent with the CLSes created from the odd-sided faces (as seen in Figure \ref{sfig:basis}(a)(iv)), so these states are already contained in our basis.
	In fact, there are no additional extended states that are linearly independent from the two cycle CLSes per unit cell, leaving $2$ flat-band eigenstates per unit cell and reflecting the two-fold degenerate gapped flat bands seen in the band spectrum (Figure \ref{sfig:D=42e2o}(d)).

	\begin{figure}[tb]
		\centering
		\includegraphics[width=\columnwidth]{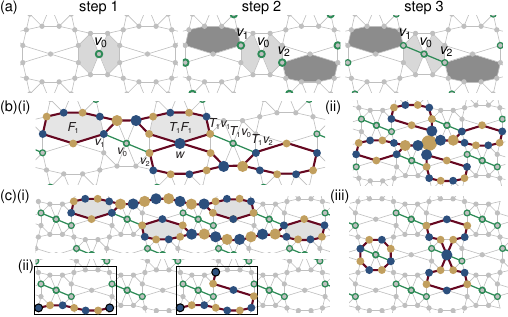}
		\refstepcounter{Sfig}\label{sfig:D=44o}
		\caption{Constructing hopping perturbations for $D=4$ line-graph lattices in the 4o family, to split the fourfold-degenerate gapped flat band at energy $-2$ and create two-fold-degenerate gapped flat bands at the same energy. \textbf{(a)} The steps of the 4o hopping perturbation construction for Example 2 of the main text, which is in the 4o family. Perturbations are in green. \textbf{(b)(i)} A compound chain CLS, whose construction is described in the main text, used to define the new flat-band basis. The wavefunction is shown as blue and yellow circles denoting real-valued relative wavefunction amplitudes of positive and negative sign, respectively. Large circles reflect twice the amplitude of smaller circles. \textbf{(ii)} A second compound chain CLS used to define the new flat-band basis. \textbf{(c)(i)} Sum of two compound chain CLSes to demonstrate how many of them can be added together to realize a superposition of two extended states. \textbf{(ii)} Two unit cells enclosed by rectangles, showing the two possible extended states. \textbf{(iii)} Two CLSes of the unperturbed lattice, which lose compact support upon perturbation, are not flat-band eigenstates of the perturbed lattice, and (with translations) span a subspace corresponding to two states per lattice site. The band spectrum for Example 2 post-perturbation can be found in Figure \ref{fig:split} of the main text.}
	\end{figure}
	
	{\bf 4o line-graph lattices:} The procedure for finding our hopping perturbation for 4o lattices is depicted in Figure \ref{sfig:D=44o}(a). We will add two new hoppings that will be $C_2$ images of one another.
	\begin{enumerate}
		\item Begin with a vertex ($v_0$) that is on a maximal Wyckoff position. We have previously shown that at least two of the four maximal Wyckoff positions are occupied by vertices, so these vertices must exist. This vertex will be at the boundary of two odd-sided faces which are $C_2$ images of each other. In Figure \ref{sfig:D=44o}(a), we pick $v_0$ to be the vertex at the boundary of both pentagon faces, shaded in light gray.
		\item Identify two odd-sided faces that are $C_2$ images of each other, have a different number of sides from the faces in step 1, and each share a vertex ($v_1$ and $v_2 = C_2 v_1$) with one of faces in step 1. These shared vertices will be part of the hopping perturbation, and must exist because the two odd-sided faces of one size must somewhere be adjacent to the two odd-sided faces of the other size, to create a connected lattice. In Figure \ref{sfig:D=44o}(a), these are the heptagon faces shaded in dark gray.
		\item Add new hoppings between vertices $v_0$ and $v_1$, as well as between $v_0$ and $v_2$. In Figure \ref{sfig:D=44o}(a), these are drawn as green edges.
	\end{enumerate} 

	After this perturbation, we find the CLS basis states to be highly nonintuitive.
	We will refer to them as ``compound chain CLSes'', as they consist of two chain-CLS-like states providing compact support for one another through the hopping perturbations, see Figure \ref{sfig:D=44o}(b)(i) for an example.
	To create these states, we begin by identifying one of the odd-sided faces from step 2 of our perturbation construction; note that this face is not adjacent to the central perturbation vertex $v_0$.
	In Figure \ref{sfig:D=44o}(b)(i), this is a heptagon face $F_1$.
	Then, we form a chain-CLS-like state that connects that face $F_1$ with an identical face translated by a single lattice vector, $T_1 F_1$.
	At the boundary of $F_1$ will be one of the perturbation vertices $v_1$ or $v_2$ identified in step 2 of the hopping construction; by forming a chain-CLS-like state with its translated copy, it will have nonzero amplitude on two copies of this perturbation vertex ($v_1$ and $T_1 v_1$ or $v_2$ and $T_1 v_2$), which are separated by the lattice translation vector $\mathbf{a}_1$.
	In Figure \ref{sfig:D=44o}(b)(i), we have nonzero amplitude on the vertices $v_1$ and $T_1 v_1$.
	Correspondingly, there are two hopping perturbation pairs associated with the two copies of this perturbation vertex: the first hopping pair consists of hopping between $v_0$ and $v_1$, and between $v_0$ and $v_2$, while the second consists of hopping between $T_1 v_0$ and $T_1 v_1$, and between $T_1 v_0$ and $T_1 v_2$.
	We next add the $C_2$ image of this chain-CLS-like state, inverting about the maximal Wyckoff position $w$ between the two central perturbation vertices $v_0$ and $T_1 v_0$.
	This provides a chain-CLS-like state with opposite amplitude on $v_2$ and $T_1 v_2$, providing compact support.
	By selecting the other lattice vector $\mathbf{a}_2$ when forming the compound chain CLS, a second flat-band eigenstate (per unit cell) can be generated, see for example the compound chain CLS in Figure \ref{sfig:D=44o}(b)(ii).
	
	Each of these two compound chain CLSes can be added to copies of itself to cover the torus and completely annihilate (analogous to property \ref{itm:FBannihil}); thus there are two linear dependencies.
	Among both compound chain CLSes, we then create $2M-2$ flat-band states for a lattice with $M$ unit cells.
	More specifically, if a compound chain CLS $|\psi\rangle$ is added to copies of itself, for all translations by the same lattice vector used in its construction (say, $T_1$), the result $\sum_i^{N_1} (T_1)^i |\psi\rangle$ is an equal superposition of a pair of extended states that wrap around the same torus direction.
	The sum of the first two terms in this summation is shown in the left of Figure \ref{sfig:D=44o}(c)(i) as an example.
	Copies of this extended-state superposition can then be translated by the other lattice vector $T_2$ and combined to yield the zero function.
	At the same time, taking just one of the extended states in this extended-state superposition, this state is linearly independent from the compound chain CLSes.
	Similarly, there is a second linearly independent extended state associated with the other compound cycle CLS.
	From inspecting the unit cell with the hopping perturbation (see Figure \ref{sfig:D=44o}(c)(ii)), there are no other extended states with compact support; all other extended states have nonzero amplitude on one of the perturbation vertices $v_0$, $v_1$, or $v_2$.
	Hence we find a total of $2M$ states.

	Finally, we must confirm that there is a subspace of dimension $2M$ whose states were eigenstates of the flat band pre-perturbation, but lose compact support upon perturbation and therefore cannot continue to be energy $-2$ flat-band eigenstates.
	For example, in Example 2 of the main text with our constructed hopping perturbation, consider a cycle and a chain CLS that are both $C_2$ eigenstates about the central perturbation vertex $v_0$, shown in Figure \ref{sfig:D=44o}(c)(iii).
	The cycle CLS has equal amplitude on the two peripheral perturbation vertices $v_1$ and $v_2$, and is therefore linearly independent from the compound chain CLSes, which both have opposite amplitude on these vertices.
	The chain CLS has nonzero amplitude on the central perturbation vertex $v_0$, and is therefore also linearly independent from the compound chain CLSes, which both have zero amplitude on this vertex.
	The cycle and chain CLS are also linearly independent from our two extended states in the post-perturbation basis, as these both have zero amplitude on all perturbation vertices.
	Therefore, the chain and cycle CLS span a subspace of dimension equal to their number: $2M$, from one chain and one cycle CLS per unit cell.
	
	More generally for the 4o lattices, we always have a cycle and a chain CLS that are (1) both flat-band eigenstates of the unperturbed lattice but not flat-band eigenstates of the perturbed lattice, (2) $C_2$ eigenstates about central perturbation vertex $v_0$, and (3) linearly independent from the perturbed flat-band basis.
	The cycle CLS encircles the two odd-sided faces in step 1 of the hopping construction (\emph{e.g.} the two pentagons in light gray of Figure \ref{sfig:D=44o}(a)).
	The chain CLS encircles two odd-sided faces that are $C_2$ images of each other and share at least one vertex with the faces in step 1, but are not the faces identified in step 2 of the construction (\emph{e.g.} the two unshaded heptagons above and below the shaded pentagons of Figure \ref{sfig:D=44o}(a)).
	As in the specific example of Figure \ref{sfig:D=44o}(c), these two CLSes for any 4o lattice will be linearly dependent from the compound chain CLSes and extended states, and span a subspace of dimension $2M$.
	
	Both of these CLSes lose compact support upon perturbation.
	Furthermore, because the cycle CLS has equal amplitude on the two peripheral perturbation vertices $v_1$ and $v_2$ (and $C_2$ eigenvalue $+1$ for inversion about the central perturbation vertex $v_0$), it cannot be combined with translated copies of itself or the chain CLS to create a state that remains an energy $-2$ eigenstate (which must have zero or opposite amplitude on $v_1$ and $v_2$ to maintain compact support).
	Because the chain CLS has zero amplitude on $v_1$ and $v_2$ and nonzero amplitude on $v_0$, it also cannot be combined with translated copies of itself or the cycle CLS to create a state that remains an energy $-2$ eigenstate (which must have zero amplitude on $v_0$ to maintain compact support).
	Thus, we find a set of states that span a subspace of dimension $2M$ and were eigenstates of the flat band pre-perturbation, but lose compact support upon perturbation and are not energy $-2$ flat-band eigenstates of the perturbed lattice.
	The energy $-2$ eigenspace is then spanned by exactly $2$ states per unit cell for these lattices after our specified hopping perturbation, yielding doubly degenerate gapped bands.
	
	We note that there exist other hopping perturbations that can split the bands in both families of $D=4$ line-graph lattices, but as these seem to depend on the particular connectivities of the specific line-graph lattices, we do not discuss them.
	
	\subsubsection{Degeneracy $D=3$}
	
	Unfortunately, for $D=3$ line-graph lattices we have not found any hopping perturbations that will split their bands as desired.
	Recall (Table \ref{stab:geoms}) that every root-graph unit cell for $D=3$ consists of two odd-sided faces and one even-sided face.
	
	\begin{figure}[tb]
		\centering
		\includegraphics[width=\columnwidth]{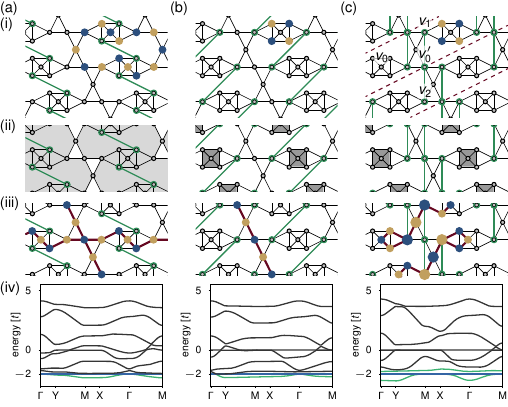}
		\refstepcounter{Sfig}\label{sfig:D=3hop}
		\caption{Example hopping perturbations (green) to the line graph of the dodecagon-triangle-triangle lattice with $D=3$ and $C_2$ symmetry. \textbf{(a)} A hopping perturbation that crosses through the even-sided face and has hopping between two vertices that are each at the boundary of the even-sided face and one odd-sided face. \textbf{(b)} A hopping perturbation that crosses through the even-sided face and has hopping between vertices that are each at the boundary of two copies of the even-sided face. \textbf{(c)} A pair of hopping perturbations with a shared vertex that lies at the boundary of two even-sided faces. In all subfigures, \textbf{(i)} shows CLSes, \textbf{(ii)} shows cycle-CLS coverings of the torus (if any), and \textbf{(iii)} shows extended states in our constructed complete basis. Blue (yellow) circles denote positive (negative) relative wavefunction amplitude, and large circles denote twice the wavefunction amplitude of smaller circles. \textbf{(iv)} Resulting band spectra, with flat bands in blue and flat bands that have been made dispersive through perturbation in green. None of these perturbations create doubly degenerate gapped flat bands.}
	\end{figure}
	
	First we consider a single hopping perturbation; in this case, the vertices involved in the hopping can no longer provide compact support and there cannot be amplitude on these vertices.
	Additionally, this perturbation must extend across an even-sided face to preserve the $C_2$ lattice symmetry, \emph{i.e.} be centered on a maximal Wyckoff position that was at the center of a face before perturbation, hence we consider perturbation vertices that are at the boundary of this face as in Figure \ref{sfig:D=3hop}(a)(i).
	If the vertices are also adjacent to the odd-sided faces, then only one CLS remains supported: the one that encircles the entire unit cell (see a(i)).
	As this covers the torus (a(ii)), one cycle CLS is linearly dependent (\ref{itm:FBannihil}).
	As two extended states are linearly independent with these states (a(iii)), we find $M+1$ states total for a lattice with $M$ unit cells, resulting a single flat band with a band touching (a(iv)).
	Therefore, the vertices cannot be adjacent to the odd-sided faces; the perturbation must only use vertices that are adjacent to two copies of the even-sided faces and are not separated by lattice translation vectors.
	
	Of the $D=3$ line graphs, we identify only the line graph of the dodecagon-triangle-triangle lattice, shown in Figures \ref{sfig:onvertexpert}(a) and \ref{sfig:D=3hop}, to contain such vertices.
	The resulting flat bands are indeed doubly degenerate (Figure \ref{sfig:D=3hop}(b)(iv)), but ungapped from a third band.
	This can also be understood by counting eigenstates (shown in (b)(i)-(iii)); we do not elaborate on the counting here because this single case can be examined directly through its band structure.
	
	We can alternatively consider adding two hopping perturbations that share a vertex that sits on a maximal Wyckoff position, as in the $D=4$ 4o hopping perturbation construction.
	If this shared vertex is adjacent to the two odd-sided faces, as in $v_0$ of Figure \ref{sfig:D=3hop}(c)(i), then the two CLSes encircling these faces (triangles in Figure \ref{sfig:D=3hop}(c)(i)) will no longer be supported and there cannot be more than one CLS per unit cell.
	Instead consider the case where this shared vertex is adjacent to two copies of the even-sided faces, as in $v'_0$ of Figure \ref{sfig:D=3hop}(c)(i).
	In this case, a set of parallel boundaries (dotted red lines in Figure \ref{sfig:D=3hop}(c)(i)) can be drawn that intersect only translated copies of this vertex $v'_0$.
	These boundaries partition the lattice into a set of ``stripes''.
	Because $v'_0$ does not have compact support, there cannot be any eigenfunction amplitude on $v'_0$ or its translated copies.
	In Figure \ref{sfig:D=3hop}(c)(i), this restricts the flat-band eigenstates to those encircling two triangle faces, such as the eigenstate shown.
	Next, considering the two other vertices $v_1$ and $v_2$ in the hopping perturbation, we find they must be located in different stripes because they must be $C_2$ images of each other for the perturbation to preserve $C_2$ symmetry.
	Then any flat-band eigenfunction of the perturbed lattice with nonzero amplitude on $v_1$ must have equal and opposite amplitude on $v_2$ to maintain compact support.
	As a result, in our example the other wavefunction encircling two triangle faces must in fact consist of the sum of its translations extending across all stripes, as shown in Figure \ref{sfig:D=3hop}(c)(iii).
	With only one CLS per unit cell, we have at most one flat band, plus band touching to a dispersive band from the extended states (Figure \ref{sfig:D=3hop}(c)(iv)).
	Thus when we add a pair of new hoppings to $D=3$ line-graph lattices, constructed in a similar way to the $D=4$ 4o hopping perturbation construction, we have at most one CLS per unit cell and at most one flat band (with possible additional degeneracies from a dispersive band).
	
	We have explored other symmetry-preserving hopping perturbations for $D=3$ lattices with $C_2$ symmetry and found that none of them create two-fold-degenerate gapped flat bands, but our search was not exhaustive.
	We also do not have a proof that such a perturbation does not exist.
	However, we note that as the number of vertices involved in the perturbations added increases, the number of vertices providing compact support decreases and fewer CLSes are supported.
	Thus beyond one or two hopping perturbations with at most three vertices involved, it is unlikely that the bands can be split.

	\subsection{Line-graph lattices with $C_3$ or $C_6$ symmetry}
	
	We claim that we cannot split the bands for line-graph lattices with $C_3$ or $C_6$ symmetry, using either on-site-energy or hopping perturbations.
	
	\begin{figure}[tb]
		\centering
		\includegraphics[width=\columnwidth]{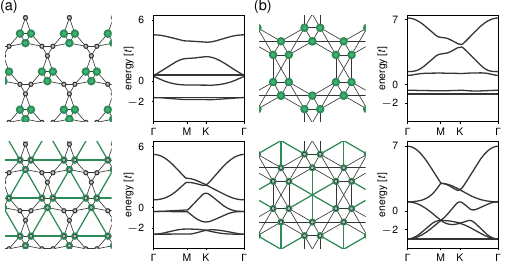}
		\refstepcounter{Sfig}\label{sfig:C3C6pert}
		\caption{The line graphs of the \textbf{(a)} nonagon-triangle lattice and the \textbf{(b)} kagome lattice with symmetry-preserving on-site-energy perturbations (top row) or hopping perturbations (bottom row), along with the resulting band spectra. None of these perturbations yield two-fold-degenerate gapped flat bands at energy $-2$. Notice also that because the line graph of the kagome lattice only has $6$ vertices per unit cell, a symmetry-preserving on-site-energy perturbations shifts all vertices by the same energy, and the resulting spectra simply shifts in energy.}
	\end{figure}
	
	For lattices with $C_3$ or $C_6$ symmetry, recall from the discussion at the beginning of this Appendix (\ref{appx:perturb}) that a minimum of one on-site energy perturbation (per unit cell) is possible if the line-graph lattice degree is a multiple of 3 or 6, respectively.
	However, for $D=3$ and $D=4$ we do not have lattices of such degree (see Table \ref{stab:geoms}), thus we need $3$ and $6$ on-site energy perturbations, which is only possible if the corresponding line-graph lattices have more than $3$ or $6$ vertices per unit cell.
	The resulting band structures can be examined directly because there are so few $D=3$ and $D=4$ line-graph lattices fulfilling these constraints, totaling $3$ in number.
	Not surprisingly, none of these perturbations split the bands to create doubly degenerate gapped bands.
	As previously noted, as the number of vertices involved in the perturbation increases, the number of vertices providing compact support decreases and fewer CLSes are supported, so with such a high number of perturbations it is unlikely that the bands will be split as desired.
	
	Similarly, for hopping perturbations a minimum of three hoppings are required.
	In this case, there are only $3$ relevant line-graph lattices, so here too they can be examined directly.
	As expected, we find that none yield the desired band splitting.
	In Figure \ref{sfig:C3C6pert} we show two line-graph lattices with symmetry-preserving on-site or hopping perturbations, and their resulting band spectra.

	\section{Split-band representation}\label{appx:splitbr}
	
	Here we identify the representation of the remaining flat bands for perturbed $1<D\leq 4$ line-graph lattices with $C_2$ symmetry and identify which hopping perturbations yield topologically non-trivial bands.
	In Tables \ref{tab:D=3splitBR} and \ref{tab:D=4splitBR} we summarize our representation findings for the flat bands in $D=3$ and $D=4$ line-graph lattices with $C_2$ symmetry, pre- and post-perturbation.
	
	\subsection{$D=3$}
	
	\begin{table*}[tb]
		\centering
		\begin{minipage}[c]{\textwidth}
			\begin{tabular}{p{50pt}@{\hspace{22pt}}p{104pt}@{\hspace{22pt}}p{100pt}@{\hspace{22pt}}p{100pt}@{\hspace{22pt}}p{58pt}}
				\hline
				&maximal Wyckoff position \newline involved in perturbation& single-band \newline representation & double-band \newline representation & double-band band topology\\
				\hline
				\multirow{4}{50pt}{$(A)_{1a}\uparrow G \,\oplus$\newline$(A)_{1b} \uparrow G \oplus (A)_{1c} \uparrow G$} & \centering $1a$ & $(A)_{1a} \uparrow G$ & $(A)_{1b}\uparrow G \oplus (A)_{1c} \uparrow G$ & trivial\\
				& \centering $1b$ & $(A)_{1b} \uparrow G$ & $(A)_{1a}\uparrow G \oplus (A)_{1c} \uparrow G$ & trivial\\
				& \centering $1c$ & $(A)_{1c} \uparrow G$ & $(A)_{1a}\uparrow G \oplus (A)_{1b} \uparrow G$ & trivial\\
				& \centering $1d$ & $(A)_{1d} \uparrow G$ & $(A)_{1a}\uparrow G \oplus (A)_{1b} \uparrow G \oplus (A)_{1c} \uparrow G \ominus (A)_{1d} \uparrow G$ & fragile\\
				\hline
				\multirow{4}{50pt}{$(A)_{1a}\uparrow G \,\oplus$\newline$(A)_{1b} \uparrow G \oplus (B)_{1c} \uparrow G$} & \centering $1a$ & $(A)_{1a} \uparrow G$ & $(A)_{1b}\uparrow G \oplus (B)_{1c} \uparrow G$ & trivial\\
				& \centering $1b$ & $(A)_{1b} \uparrow G$ & $(A)_{1a}\uparrow G \oplus (B)_{1c} \uparrow G$ & trivial\\
				& \centering $1c$ & $(B)_{1c} \uparrow G$ & $(A)_{1a}\uparrow G \oplus (A)_{1b} \uparrow G$ & trivial\\
				& \centering $1d$ & $(B)_{1d} \uparrow G$ & $(A)_{1a}\uparrow G \oplus (A)_{1b} \uparrow G \oplus (B)_{1c} \uparrow G \ominus (B)_{1d} \uparrow G$ & fragile\\
				\hline
			\end{tabular}
			\refstepcounter{Stab}\label{tab:D=3splitBR}
			\caption{Summary of representations for $D=3$, $C_2$-symmetric line-graph lattices. The first column is the representation of the flat bands pre-perturbation. For a hopping perturbation that involves a graph element sitting on the maximal Wyckoff position in the second column, the resulting single-band and double-band representations are listed, along with the double-band band topology. Notice that we make some choices without loss of generality: in the first row, the $1a$, $1b$, and $1c$ positions are occupied by vertices, and $1d$ is occupied by a face; in the second, $1a$ and $1b$ are occupied by vertices, $1c$ by a complete subgraph, and $1d$ by a face.}
		\end{minipage}
	\end{table*}

	For perturbed $D=3$ line-graph lattices with $C_2$ symmetry, considering the two ungapped split bands with one flat and one dispersive, they will exhibit fragile topology if the perturbation crosses through the center of an even-sided face that is on a maximal Wyckoff position.
	
	From Table \ref{table:maxWyckpos} in the main text, we see that for $D=3$ line-graph lattices with $C_2$ symmetry, the four RSIs (one for each maximal Wyckoff position) will be $0$, $-1$, $-1$, and $-1$ if the root graph degree is odd, and $0$, $-1$, $-1$, and $+1$ if the degree is even.
	Without loss of generality, take the maximal Wyckoff positions occupied by faces and (if applicable) complete subgraphs to be the $1d$ and $1c$ positions.
	Then the respective representations are $(A)_{1a} \!\uparrow\! G \oplus (A)_{1b} \!\uparrow\! G \oplus (A)_{1c}\!\uparrow\! G$ and $(A)_{1a} \!\uparrow\! G \oplus (A)_{1b} \!\uparrow\! G \oplus (B)_{1c} \!\uparrow\! G$.
	
	Unfortunately, because one of the ungapped bands will be dispersive after perturbation, we cannot use our developed formalism to examine the $C_2$ eigenfunctions and determine the new RSIs and representations.
	However, we can directly compute the representations \cite{Bradlyn2017s}.
	From this we find evidence that if the perturbation sits on the $1a$, $1b$, or $1c$ positions, the band with the corresponding band representation is lifted and the remaining bands are still represented by a sum of EBRs.
	If the perturbation instead sits on the $1d$ position, then for odd root-graph degree the resulting representation becomes $(A)_{1a} \!\uparrow\! G \oplus (A)_{1b} \!\uparrow\! G \oplus (A)_{1c} \!\uparrow\! G \ominus (A)_{1d} \!\uparrow\! G$, and for even root-graph degree it is $(A)_{1a} \!\uparrow\! G \oplus (A)_{1b} \!\uparrow\! G \oplus (B)_{1c} \!\uparrow\! G \ominus (B)_{1d} \!\uparrow\! G$.
	Although these decomposition are not unique, all such decompositions give fragile topology.
	
	\subsection{$D=4$}
	
	For $D=4$ lattices in the 4o family, the flat bands of its line graph will exhibit fragile topology once the bands are split according to our 4o hopping perturbation construction.
	For the 2e2o family, however, the flat bands will continue to admit a Wannier representation once the bands are split according to our 2e2o hopping perturbation construction.
	We have also considered other perturbations which split the $D=4$ band degeneracy and yield two-fold-degenerate gapped flat bands, and find the same band topology for lattices within these families (fragile topology in 4o lattices, but not for 2e2o lattices).
	However, this search is not exhaustive and we do not have a proof which considers all possible perturbations.
	In this Appendix, we determine the two-fold-degenerate flat-band representation for our constructed perturbations in the 2e2o and 4o families. 
	
	In each case, we begin by considering the graph-element types on the maximal Wyckoff positions.
	Next we consider the associated RSIs and flat-band representations of the fourfold degenerate flat bands.
	Finally, we examine how the RSIs change after perturbation to determine the  representations of the split bands.
	
	\begin{figure}
		\centering
		\includegraphics[width=\columnwidth]{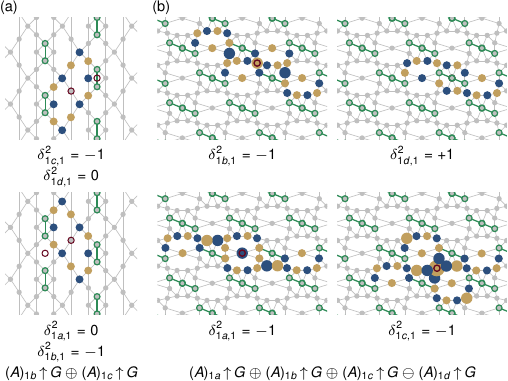}
		\refstepcounter{Sfig}\label{sfig:splitbandrep}
		\caption{$C_2$ flat-band eigenfunctions for $C_2$ centers at each of the maximal Wyckoff positions post-perturbation, in \textbf{(a)} the heptagon-heptagon-hexagon-square kagome lattice (2e2o family) and \textbf{(b)} Example 2 (4o family). Hopping perturbations are shown in green, maximal Wyckoff positions as red outlined circles, and wavefunctions as blue and yellow circles denoting relative positive and negative real-valued wavefunction amplitude, respectively, with large circles having twice the amplitude of smaller ones. Notice that in (a), the vertex-occupied maximal Wyckoff positions are labeled as $1b$ and $1c$.}
	\end{figure}
	
	\subsubsection{2e2o family}
	
	If two faces of the root-graph unit cell are odd-sided and two are even-sided, then two of the four maximal Wyckoff positions are occupied by vertices and two are occupied by faces in the line graph.
	This is because even-sided faces in the root graph must sit on maximal Wyckoff positions (\ref{itm:RGface}), of which two are present, and there must be at least two maximal Wyckoff positions occupied by edges in the root graph (\ref{itm:RGedges}).
	
	Subsequently (by \ref{itm:RSI}) we have RSIs $\delta^2_{w', 1} = 0$ on the two maximal Wyckoff positions $w'$ occupied by faces, and $\delta^2_{w'', 1} = -1$ on the two maximal Wyckoff positions $w''$ occupied by vertices.
	Without loss of generality, take the maximal Wyckoff positions occupied by vertices to be the $1a$ and $1b$ positions; then the representation can be chosen as $(A)_{1a} \!\uparrow\! G \oplus (A)_{1b} \!\uparrow\! G \oplus (A)_{w} \!\uparrow\! G \oplus (B)_{w} \!\uparrow\! G$, where $w$ can be $1a, 1b, 1c,$ or $1d$.
	From examining the basis states after adding a single hopping across one of the even-sided faces (done in Appendix \ref{appx:perturb}), we see in Figure \ref{sfig:splitbandrep}(a) that the RSIs  $\delta^2_{w', 1} = 0$ and  $\delta^2_{w'', 1} = -1$ remain invariant post-perturbation.
	However, because there are now a total of two flat bands, the representation of these bands becomes $(A)_{1a} \!\uparrow\! G \oplus (A)_{1b} \!\uparrow\! G$.
	As this is a sum of EBRs, these bands may be topologically trivial.
	
	We do not consider perturbations for this family of line-graph lattices involving the vertices sitting on maximal Wyckoff positions, however we conjecture that a hopping that intersects the $1a$ Wyckoff position will split the bands into doubly degenerate gapped bands with a representation of $(B)_{1a} \!\uparrow\! G \oplus (A)_{1b} \!\uparrow\! G$.
	
	\subsubsection{4o family}
	
	\begin{table*}[tb]
		\centering
		\begin{minipage}[c]{\textwidth}
			\begin{tabular}{p{10pt}p{40pt}@{\hspace{22pt}}p{104pt}@{\hspace{22pt}}p{100pt}@{\hspace{22pt}}p{100pt}@{\hspace{22pt}}p{58pt}}
				\hline
				&&maximal Wyckoff position \newline involved in perturbation& both single-band \newline representations & double-band \newline representation & double-band band topology\\
				\hline
				\multirow{18}{*}{\rotatebox{90}{4o}}&\multirow{4}{50pt}{$(A)_{1a}\uparrow G \,\oplus$\newline$(A)_{1b} \uparrow G \,\oplus$\newline$(A)_{1c} \uparrow G \oplus (A)_{1d} \uparrow G$} & \centering $1a$ & $(A)_{1a} \uparrow G$ \& $(A)_{1a} \uparrow G$ & $\ominus (A)_{1a}\uparrow G \oplus (A)_{1b}\uparrow G \oplus (A)_{1c} \uparrow  G \oplus (A)_{1d}\uparrow G$ & fragile\\
				&& \centering $1b$ & $(A)_{1b} \uparrow G$ \& $(A)_{1b} \uparrow G$ & $(A)_{1a}\uparrow G \ominus (A)_{1b}\uparrow G \oplus (A)_{1c} \uparrow  G \oplus (A)_{1d}\uparrow G$ & fragile\\
				&& \centering $1c$ & $(A)_{1c} \uparrow G$ \& $(A)_{1c} \uparrow G$ & $(A)_{1a}\uparrow G \oplus (A)_{1b}\uparrow G \ominus (A)_{1c} \uparrow  G \oplus (A)_{1d}\uparrow G$ & fragile\\
				&& \centering $1d$ & $(A)_{1d} \uparrow G$ \& $(A)_{1d} \uparrow G$ & $(A)_{1a}\uparrow G \oplus (A)_{1b}\uparrow G \oplus (A)_{1c} \uparrow  G \ominus (A)_{1d}\uparrow G$ & fragile\\
				\cline{2-6}
				&\multirow{4}{50pt}{$(A)_{1a}\uparrow G \,\oplus$\newline$(A)_{1b} \uparrow G \,\oplus$\newline$(B)_{1c} \uparrow G \oplus (B)_{1d} \uparrow G$} & \centering $1a$ & $(A)_{1a} \uparrow G$ \& $(A)_{1a} \uparrow G$ & $\ominus (A)_{1a}\uparrow G \oplus (A)_{1b}\uparrow G \oplus (B)_{1c} \uparrow  G \oplus (B)_{1d}\uparrow G$ & fragile\\
				&& \centering $1b$ & $(A)_{1b} \uparrow G$ \& $(A)_{1b} \uparrow G$ & $(A)_{1a}\uparrow G \ominus (A)_{1b}\uparrow G \oplus (B)_{1c} \uparrow  G \oplus (B)_{1d}\uparrow G$ & fragile\\
				&& \centering $1c$ & $(B)_{1c} \uparrow G$ \& $(B)_{1c} \uparrow G$ & $(A)_{1a}\uparrow G \oplus (A)_{1b}\uparrow G \ominus (B)_{1c} \uparrow  G \oplus (B)_{1d}\uparrow G$ & fragile\\
				&& \centering $1d$ & $(B)_{1d} \uparrow G$ \& $(B)_{1d} \uparrow G$ & $(A)_{1a}\uparrow G \oplus (A)_{1b}\uparrow G \oplus (B)_{1c} \uparrow  G \ominus (B)_{1d}\uparrow G$ & fragile\\
				\hline
				\multirow{4}{10pt}{\rotatebox{90}{2e2o}}&\multirow{4}{50pt}{$(A)_{1a}\uparrow G \,\oplus$\newline$(A)_{1b} \uparrow G \,\oplus$\newline$(A)_{w} \uparrow G \oplus (B)_{w} \uparrow G$} & \centering $1a$ & $(A)_{1a} \uparrow G$ \& $(A)_{1a} \uparrow G$& $(B)_{1a}\uparrow G \oplus (A)_{1b}\uparrow G$ & trivial\\
				&& \centering $1b$ & $(A)_{1b} \uparrow G$ \& $(A)_{1b} \uparrow G$ & $(A)_{1a}\uparrow G \oplus (B)_{1b} \uparrow G$ & trivial\\
				&& \centering $1c$ & $(A)_{1c} \uparrow G$  \& $(B)_{1c} \uparrow G$ & $(A)_{1a}\uparrow G \oplus (A)_{1b} \uparrow G$ & trivial\\
				&& \centering $1d$ & $(A)_{1d} \uparrow G$ \& $(B)_{1d} \uparrow G$ & $(A)_{1a}\uparrow G \oplus (A)_{1b} \uparrow G$ & trivial\\
				\hline
			\end{tabular}
			\refstepcounter{Stab}\label{tab:D=4splitBR}
			\caption{Summary of representations for $D=4$, $C_2$-symmetric line-graph lattices. The first column is the representation of the flat bands pre-perturbation. For a hopping perturbation that involves a graph element sitting on the maximal Wyckoff position in the second column, the resulting two single-band and the double-band representations are listed, along with the double-band band topology. Notice that we make some choices without loss of generality: in the first row, the $1a$, $1b$, $1c$, and $1d$ positions are occupied by vertices; in the second, $1a$ and $1b$ are occupied by a vertex, while $1c$ and $1d$ are occupied by complete subgraphs; in the third, $1a$ and $1b$ are occupied by vertices and $1c$ and $1d$ by faces. We find that we can split the bands in all 4o line-graph lattices to yield fragile topological bands, but get topologically trivial bands in the 2e2o lattices.}
		\end{minipage}
	\end{table*}
	
	If all four faces of the root-graph unit cell are odd-sided and the root-graph lattice is of odd degree, all four maximal Wyckoff positions are occupied by edges in the root graph, \textit{i.e.} vertices in the line graph.
	This is because maximal Wyckoff positions cannot be occupied by odd-sided faces or on vertices of odd degree.
	
	If instead the root graph is of even degree, vertices sit on either two or four maximal Wyckoff positions in the line graph and the remaining are occupied by complete subgraphs.
	This is because maximal Wyckoff positions cannot be occupied by odd-sided faces, there must be an even number of maximal Wyckoff positions occupied by vertices in the line graph (\emph{i.e.} edges in the root graph) by \ref{itm:RGparity}, and there must be edges on at least two maximal Wyckoff positions in the root graph by \ref{itm:RGedges}.
	
	We have RSIs of $\delta^2_{w', 1} = -1$ on the vertex-occupied maximal Wyckoff positions $w'$ and $\delta^2_{w'', 1} = +1$ on the maximal Wyckoff positions $w''$ occupied by complete subgraphs.
	Without loss of generality, label two vertex-occupied maximal Wyckoff positions as the $1a$ and $1b$ positions.
	The resulting representation is then $(A)_{1a} \!\uparrow\! G \oplus (A)_{1b} \!\uparrow\! G \oplus (A)_{1c} \!\uparrow\! G \oplus (A)_{1d} \!\uparrow\! G$ (for four vertex maximal Wyckoff positions) or $(A)_{1a} \!\uparrow\! G \oplus (A)_{1b} \!\uparrow\! G \oplus (B)_{1c} \!\uparrow\! G \oplus (B)_{1d} \!\uparrow\! G$ (for two maximal Wyckoff positions occupied by vertices and two occupied by complete subgraphs).
	
	Again without loss of generality, pick the $1a$ maximal Wyckoff position as the position occupied by the central perturbation vertex for the hopping pair (defined as $v_0$ in Appendix \ref{appx:perturb}2a).
	After this perturbation, our constructed basis states include compound chain CLSes, see Appendix \ref{appx:perturb}2a.
	Even so, as seen in Figure \ref{sfig:splitbandrep}(b) we can use them to determine the post-perturbation RSIs.
	For each of the three maximal Wyckoff positions on unperturbed graph elements, a compound chain CLS centered about that position exists, with the same $C_2$ eigenvalue as the pre-perturbation $C_2$ eigenstate.
	The $1a$ $C_2$ center $C_2$ eigenfunction, by contrast, is a cycle CLS encircling all four faces of the unit cell and has eigenvalue $+1$.
	As before, all other $C_2$ eigenstate constructions for each maximal Wyckoff position can be paired as CLSes plus or minus their $C_2$ images, such that they do not affect the RSI.
	The new RSIs are then $-1$, $-1$, $-1$, $+1$ for four vertex-occupied maximal Wyckoff positions and $-1$, $+1$, $+1$, $+1$ for two vertex- and two complete-subgraph-occupied maximal Wyckoff positions.
	The resulting doubly degenerate flat-band representations can be chosen as $\ominus(A)_{1a} \!\uparrow\! G \oplus (A)_{1b} \!\uparrow\! G \oplus (A)_{1c} \!\uparrow\! G \oplus (A)_{1d} \!\uparrow\! G$ and $\ominus(A)_{1a} \!\uparrow\! G \oplus (A)_{1b} \!\uparrow\! G \oplus (B)_{1c} \!\uparrow\! G \oplus (B)_{1d} \!\uparrow\! G$, respectively, indicating fragile topological bands in both cases.
	Note that while this decomposition is not unique, all such decompositions give fragile topology.

	We do not consider here perturbations involving complete subgraphs that sit on maximal Wyckoff positions.
	However, we conjecture that a hopping that intersects the $1c$ Wyckoff position will split bands with a representation $(A)_{1a} \!\uparrow\! G \oplus (A)_{1b} \!\uparrow\! G \oplus (B)_{1c} \!\uparrow\! G \oplus (B)_{1d}$ and create doubly degenerate gapped bands with a representation of $(A)_{1a} \!\uparrow\! G \oplus (A)_{1b} \!\uparrow\! G \ominus (B)_{1c} \!\uparrow\! G \oplus (B)_{1d} \!\uparrow\! G$.

	\section{Additional examples}\label{appx:exs}
	
	\begin{figure*}[t]
		\centering
		\begin{minipage}[c]{\textwidth}
			\includegraphics[width=0.95\textwidth]{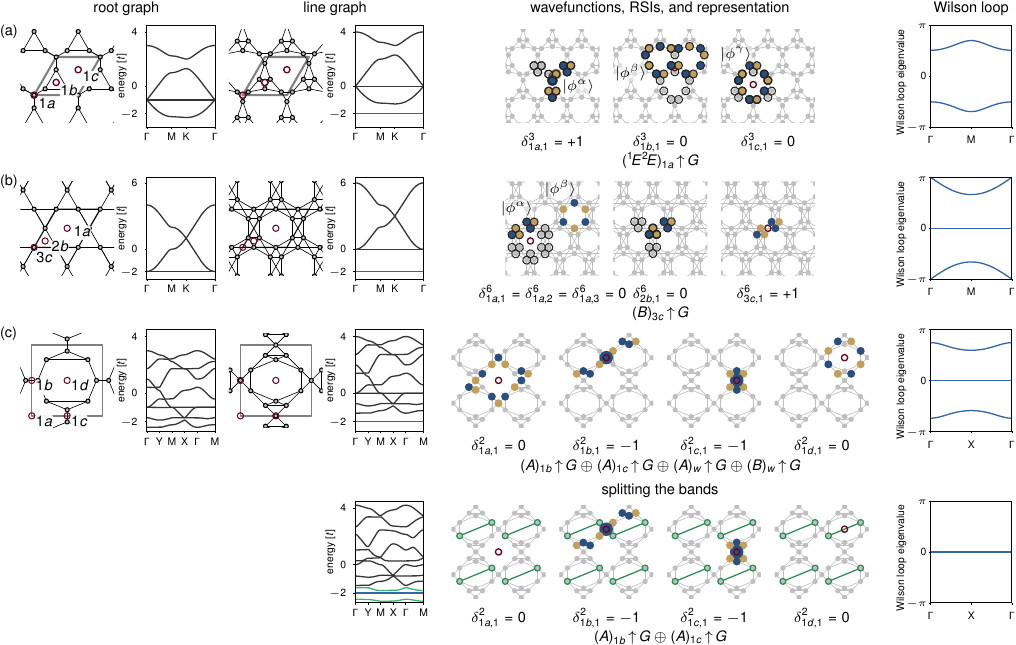}
			\refstepcounter{Sfig}\label{sfig:additionalexs}
			\caption{Additional examples. For each lattice, we show (from left to right) the root graph; root-graph spectrum; line graph; line-graph spectrum; construction of symmetry and energy eigenfunctions with resulting RSIs and representation; and Wilson loop winding. \textbf{(a)} Line graph of the nonagon-triangle lattice. \textbf{(b)} Line graph of the kagome lattice. \textbf{(c)} Line graph of the dodecagon-hexagon-triangle-triangle lattice. For this last lattice, we also show a perturbation that successfully splits the bands, in addition to the resulting spectrum, RSIs, representation, and Wilson loop winding. Grey outlined parallelograms show a single unit cell, red outlined circles denote maximal Wyckoff positions, blue bands in the band spectra highlight flat bands, and green bands in the band spectra highlight flat bands that have been perturbed to become dispersive. Lattice perturbations are also shown in green, and blue and yellow circles represent energy eigenfunctions, with blue (yellow) denoting relative positive (negative) wavefunction amplitude and large circles denoting twice the amplitude of smaller circles.}
		\end{minipage}
	\end{figure*}

	To further demonstrate our developed formalism, we determine the flat-band  representation for three additional lattices: the nonagon-triangle kagome lattice ($D=2$, $C_3$ symmetry), the line graph of the kagome lattice ($D=3$, $C_6$ symmetry), and the dodecagon-hexagon-triangle-triangle kagome lattice ($D=4$, $C_2$ symmetry).
	All three are shown in Figure \ref{sfig:additionalexs}.
	
	\textbf{Line graph of nonagon-triangle:} The line graph of the nonagon-triangle lattice is the only other $D=2$ line-graph lattice in our set besides the line graph of the triangle lattice.
	It has $d=3$, and $n=4$.
	According to Table \ref{table:maxWyckpos} of the main text and as seen from the red empty circles in the first column of Figure \ref{sfig:additionalexs}(a), of its three maximal Wyckoff positions, the root-graph lattice has faces on two positions and a vertex on one position.
	
	To determine the RSIs at each of these maximal Wyckoff positions, we identify three linearly independent local energy eigenfunctions $|\phi^\alpha\rangle$, $|\phi^\beta\rangle$, and $|\phi^\gamma\rangle$ as shown in Figure \ref{sfig:additionalexs}(a).
	These can be used to construct the $C_3$ eigenfunctions at each of the Wyckoff positions.
	We now consider each of these positions in turn.
	
	For the $1a$ position, notice that of the $C_3$ eigenfunctions $|\phi_k\rangle \equiv |\phi\rangle + e^{i2\pi k/s}C_3|\phi\rangle + (e^{i2\pi k/s}C_3)^2|\phi\rangle$, when $k=0$ we have $|\phi^\alpha_0\rangle = |\phi^\alpha\rangle + C_3 |\phi^\alpha\rangle + C_3^2 |\phi^\alpha\rangle$, which vanishes completely, whereas when $k=1$, we have $|\phi^\alpha_1\rangle = |\phi^\alpha\rangle + e^{i 2\pi/3} C_3 |\phi^\alpha\rangle + (e^{i 2\pi/3} C_3)^2 |\phi^\alpha\rangle$, which does not.
	The analogous eigenfunction constructions for $|\phi^\beta\rangle$ and $|\phi^\gamma\rangle$ yield valid (non-zero) wavefunctions for both $k=0$ and $k=1$; this is the case for all other wavefunctions as well.
	Therefore, we find that $\delta^3_{1a, 1} \equiv m^3_{1a, 1} - m^3_{a, 0} = +1$.
	For the $1b$ and $1c$ positions, all eigenfunction constructions yield valid wavefunctions for both $k=0$ and $k=1$, therefore 
	$\delta^3_{1b, 1} \equiv m^3_{1b, 1} - m^3_{1b, 0} = 0$ and $\delta^3_{1c, 1} \equiv m^3_{1c, 1} - m^3_{1c, 0} = 0$.
	Given that a complete subgraph of the line-graph lattice sits at the $1a$ position while faces sit at the $1b$ and $1c$ positions, these RSIs are consistent with the relationships described in the main text: maximal Wyckoff positions $w$ with $C_3$ symmetry have RSI $\delta^3_{w, 1} = +1$ if occupied by a complete subgraph, otherwise $\delta^3_{w, 1} = 0$.
	
	The resulting representation is then $(^1\!E^2\!E)_{1a} \!\uparrow\! G$, a two-dimensional irrep with $e^{\pm i 2\pi/3}$ eigenvalues degenerate by TRS, and the bands admit a Wannier representation.
	Correspondingly, we find no Wilson loop winding.
	
	\textbf{Line graph of kagome:} Here, $d=4$ and $n=3$, and the root-graph lattice has two faces and one vertex on the three maximal Wyckoff positions.
	The local energy eigenfunctions to consider are $|\phi^\alpha\rangle$ and $|\phi^\beta\rangle$ as shown in Figure \ref{sfig:additionalexs}(b); they are linearly independent and, with lattice translations, span the entire flat-band basis.
	
	At the $1a$ position, using $|\phi^\alpha\rangle$ we find that each value of $k\in[0, 3]$ gives one $C_6$ energy eigenfunction.
	In fact, the $k=3$ eigenfunction is equal to $|\phi^\beta\rangle$, so we find an equal number of eigenfunctions of each eigenvalue, thus $\delta^6_{1a, 1} = \delta^6_{1a, 2} =\delta^6_{1a, 3} = 0$.
	Moving on to the $2b$ position, we find both $|\phi^\alpha\rangle$ and $|\phi^\beta\rangle$ construct valid wavefunctions for $k=0$ and $k=1$; $\delta^6_{2b, 1} = 0$.
	Finally, for the $3c$ position, we see that $|\phi^\alpha\rangle$ is itself a $C_2$ eigenstate of eigenvalue $-1$, while $|\phi^\beta\rangle$ can be used to produce one eigenstate of each eigenvalue.
	As a result, we find $\delta^6_{3c, 1} = +1$.
	These, too, are consistent with the relationships described in the main text: ($1a$) maximal Wyckoff positions with $C_6$ symmetry have RSIs $\delta^6_{1a, 1} = \delta^6_{1a, 2} =\delta^6_{1a, 3}$, equal to $0$ if it is occupied by a face; maximal Wyckoff positions with $C_3$ symmetry have RSI equal to $0$ if occupied by a complete subgraph, and maximal Wyckoff positions with $C_2$ symmetry have RSI equal to $+1$ if occupied by a vertex.
	
	The representation satisfying all constraints is $(B)_{3c} \!\uparrow\! G$, accompanied by no Wilson loop winding.
	
	\textbf{Line graph of dodecagon-hexagon-triangle-triangle:} In this lattice, $d=3$ and $n=8$, for a total of $D=4$ flat bands at energy $-2$.
	The root-graph unit cell has $C_2$ symmetry and two edges and two faces on the four maximal Wyckoff positions, in agreement with Table \ref{table:maxWyckpos} of the main text.
	Furthermore, following the patterns described in the main text, it has an even number of edges per unit cell, so there must be edges on an even number of maximal Wyckoff positions and edges on at least two maximal Wyckoff positions.
	When taking the line graph, these graph elements on the maximal Wyckoff positions transform into two vertices and two faces.
	
	As for the RSIs of this line-graph lattice, recall from \ref{itm:RSI} that we expect to find $\delta^2_{w, 1} = 0$ for maximal Wyckoff positions $w$ occupied by faces and $\delta^2_{w, 1} = -1$ for $w$ occupied by vertices.
	Indeed, this is the case here; for the $1a$ and $1d$ positions as shown in Figure \ref{sfig:additionalexs}, which are occupied by faces, we can construct both an even and an odd $C_2$ energy eigenfunction, with all other constructions producing an equal number of $+1$ and $-1$ $C_2$ eigenstates, as shown in Figure \ref{sfig:additionalexs}(c).
	Similarly, for the $1b$ and $1c$ positions, both occupied by vertices, we find an excess of a single $+1$ $C_2$ energy eigenfunction.
	
	This yields $(A)_{1b} \!\uparrow\! G \oplus (A)_{1c} \!\uparrow\! G \oplus (A)_{w} \!\uparrow\! G \oplus (B)_{w} \!\uparrow\! G$ for the representation, where $w$ can be any of $1a$, $1b$, $1c$, or $1d$ (since $(A+B)_{1a} \!\uparrow\! G = (A+B)_{1b} \!\uparrow\! G = (A+B)_{1c} \!\uparrow\! G = (A+B)_{1d} \!\uparrow\! G$).
	As expected, there is no Wilson loop winding.
	
	Because this line-graph lattice has $C_2$-symmetry and $D=4$, we can add a perturbation to split its bands.
	More specifically, as its root-graph unit cell is comprised of a dodecagon, hexagon, and two triangles, it falls in the 2e2o family.
	Then, as shown in Appendix \ref{appx:perturb}, we can split its bands by adding a hopping perturbation between the two vertices in the line-graph lattice sitting between the dodecagon and hexagon, drawn in green in Figure \ref{sfig:additionalexs}(c).
	Indeed, the bands are split to create twofold-degenerate gapped flat bands at energy $-2$ and two dispersive bands.
	
	Now we can re-evaluate the RSIs at each maximal Wyckoff position to determine the representation of the two flat bands.
	Upon revisiting the four maximal Wyckoff positions, we find that $1a$ and $1d$ still have an equal number of odd and even $C_2$ eigenfunctions, hence $\delta^2_{1a} = \delta^2_{1d} = 0$.
	Similarly, the $1b$ and $1c$ positions have the same extra even eigenfunction as before, so $\delta^2_{1b} = \delta^2_{1c} = -1$.
	With these same RSIs now describing a total of two bands, the resulting representation is $(A)_{1b} \!\uparrow\! G \oplus (A)_{1c} \!\uparrow\! G$.
	As predicted, and consistent with the Wilson loop winding result, it admits a Wannier representation.

	\section{Extensions of our work}\label{appx:extensions}
	
	Here we provide evidence that the formalism presented in this work can be straightforwardly extended to other line-graph lattices of regular root-graph lattices.
	
	\subsection{$D > 4$ line-graph lattices with $C_2$ symmetry}
	
\begin{figure}[t]
		\centering
		\includegraphics{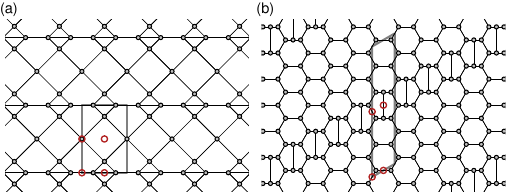}
		\refstepcounter{Sfig}\label{fig:D>4}
		\caption{\textbf{(a)} The pentagon-pentagon-square-triangle-triangle root-graph lattice with $D=5$ and $C_2$ symmetry. \textbf{(b)} The heptagon-heptagon-hexagon-hexagon-pentagon-pentagon root-graph lattice with $D=6$ and $C_2$ symmetry. Vertices are indicated as grey circles for a single unit cell, enclosed by the thick grey line. Maximal Wyckoff positions are circled in red.}
	\end{figure}

	Consider non-bipartite planar root-graph lattices with $C_2$ symmetry corresponding to line-graph lattices with $D > 4$.
	We examine several such lattices, two of which are shown in Figure \ref{fig:D>4}, and find that the relationship between graph-element type on the maximal Wyckoff positions and RSI for these lattices is the same as for $C_2$-symmetric line-graph lattices with $D \leq 4$: $\delta^2_{w, 1}$ is equal to $-1$ for maximal Wyckoff positions $w$ occupied by a vertex, $+1$ for positions occupied by complete subgraphs, and $0$ for positions occupied by faces.
	This should not be surprising given the local nature of the eigenfunctions involved in determining the RSI.
	
	Notice, however, that not all of the theorems of Appendix \ref{appx:roottolinebr} hold for these $D>4$ lattices.
	For example, the heptagon-heptagon-hexagon-hexagon-pentagon-pentagon kagome lattice of Figure \ref{fig:D>4}(b) has point-group symmetry $C_2$, and there are two faces whose number of sides is a multiple of $2$, but they do not sit on maximal Wyckoff positions.
	Instead, we conjecture that for $D$ even, there are faces on an even number of maximal Wyckoff positions, while for $D$ odd, there are faces on an odd number of such positions.
	
	If this conjecture holds, the representation of the $D$ flat bands may be generalized.
	Define $W_e$ to the be set of maximal Wyckoff positions occupied by an edge (site) of the root-graph (line-graph) lattice and $W_v$ to be the set of maximal Wyckoff positions occupied by a vertex (complete subgraph) of the root-graph (line-graph) lattice.
	Then the general representation will be given by:
	\begin{multline}
	\bigoplus_{i \in W_e} (A)_{i} \!\uparrow\! G \,\oplus\, \bigoplus_{i \in W_v} (B)_{i} \!\uparrow\! G \,\oplus\,\\ \frac{D - |W_e| - |W_v|}{2} \big((A)_{w} \!\uparrow\! G \oplus (B)_{w} \!\uparrow\! G\big)
	\end{multline}
	where $w$ can be $1a$, $1b$, $1c$, or $1d$.
	This representation is always a sum of EBRs.
	
	According to this generalization, the representation of the line-graph lattice corresponding to the root-graph lattice of Figure \ref{fig:D>4}(a) should be $(A)_{1a} \!\uparrow\! G \oplus (B)_{1b} \!\uparrow\! G \oplus (A)_{1c} \!\uparrow\! G \oplus (A)_w \!\uparrow\! G \oplus (B)_w \!\uparrow\! G$, for $w = 1a$, $1b$, $1c$, or $1d$.
	Likewise, the line-graph lattice of Figure \ref{fig:D>4}(b) should have a flat-band representation of $(A)_{1a} \!\uparrow\! G \oplus (A)_{1b} \!\uparrow\! G \oplus (A)_{1c} \!\uparrow\! G \oplus (A)_{1d} \!\uparrow\! G \oplus (A)_w \!\uparrow\! G \oplus (B)_w \!\uparrow\! G$, for $w = 1a$, $1b$, $1c$, or $1d$.
	Indeed, direct computation of the flat-band representations confirms that this is the case for both lattices.
	
	In addition to proving these claims, it remains to be shown how bands that do not admit a Wannier representation can be achieved through adding (symmetry-preserving) perturbations to these line-graph lattices.

	\subsection{Line-graph lattices with $C_4$ symmetry}
	
	\begin{figure}[t]
		\centering
		\includegraphics{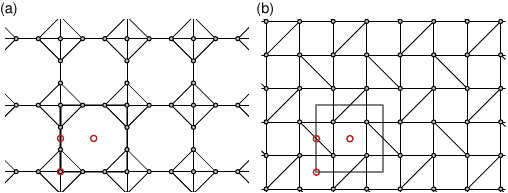}
		\refstepcounter{Sfig}\label{fig:C4}
		\caption{\textbf{(a)} The octagon-triangle-triangle-triangle-triangle root-graph lattice with $D=5$ and $C_4$ symmetry. \textbf{(b)} The square-square-triangle-triangle-triangle-triangle root-graph lattice with $D=6$ and $C_4$ symmetry. Vertices are indicated as grey circles for a single unit cell, enclosed by the thick grey line. Maximal Wyckoff positions are circled in red.}
	\end{figure}
	
	The remaining point-group symmetry to be considered for periodic lattices is $C_4$ symmetry.
	In general, these lattices must be comprised of faces that are $C_4$-symmetric and/or sets of four faces that are $C_4$ images of each other.
	As a result, none of the lattices considered in this work exhibit $C_4$ symmetry; with $D \leq 4$, the square lattice is the only such root-graph lattice, and is bipartite.
	
	However, by moving to $D > 4$, such symmetries become possible.
	We show two root-graph counterparts of these lattices in Figure \ref{fig:C4}, one for $D=5$ and one for $D=6$.
	
	The maximal Wyckoff positions for $C_4$-symmetric lattices are the $1a$ and $1b$ positions, each with $C_4$ symmetry, and the $2c$ position with $C_2$ symmetry.
	Aside from $d=4$ root-graph lattices, where the $1a$ and $2b$ positions can be occupied by vertices, the $1a$ and $2b$ positions will generally be occupied by the $C_4$ symmetric faces.
	Similarly, aside from a few specific cases, the $2c$ positions will be occupied by edges.
	
	The same approach used in determining the RSIs for $D \leq 4$ line-graph lattices can be applied here.
	For the $2c$ positions, the arguments presented in Appendix \ref{appx:roottolinebr} and Figure \ref{sfig:C2RSIs} regarding $C_2$ flat-band eigenstate constructions and the counting of $+1$ and $-1$ eigenstates still apply.
	Therefore, we see no reason why the element-type-to-RSI relationship here should be different than for $C_2$-center maximal Wyckoff positions in lattices with other symmetries.
	
	For the $1a$ and $1b$ positions, we claim that for position $w \in {1a, 1b}$ at the center of a face in the root-graph lattice, the RSIs for the corresponding line-graph lattice are $\delta^4_{w, 1} \equiv m^4_{w, 1} - m^4_{w, 0} = 0$ and $\delta^4_{w, 2} \equiv m^4_{w, 2} - m^4_{w, 0} = 0$.
	This can be seen by considering flat-band CLSes $|\phi\rangle$ encircling two odd-sided faces locally around $w$.
	Using these eigenstates to construct the $C_4$ flat-band eigenfunctions as in Equation \ref{eq:CsEigenfn}, an equal number of eigenfunctions of each eigenvalue will be generated.
	As for position $w \in {1a, 1b}$ occupied by a vertex in the root-graph lattice, the corresponding line-graph lattice has RSIs $\delta^4_{w, 1} = \delta^4_{w, 2} = 1$.
	Here, too, we can consider flat-band CLSes $|\phi\rangle$ encircling two odd-sided faces locally around $w$.
	However, now when we construct the $C_4$ flat-band eigenfunctions using Equation \ref{eq:CsEigenfn}, the eigenfunction $|\phi_0\rangle$ will be equal to the zero function, thus there will be one fewer eigenfunction of $C_4$ eigenvalue $+1$ relative to those for the other eigenvalues.
	
	This results in the correct representation for our two examples.
	The octagon-triangle-triangle-triangle-triangle lattice has the $1a$ position occupied by a vertex and $1b$ occupied by a face; thus the line-graph lattice RSIs are $\delta^4_{1a, 1} = \delta^4_{1a, 2} = 1$ and $\delta^4_{1b, 1} = \delta^4_{1b, 2} = 0$.
	Its $2c$ position is occupied by an edge, yielding $\delta^2_{2c, 1} = -1$.
	The corresponding representation for the five-fold-degenerate gapped flat bands in the line-graph lattice is $({}^1\!E{}^2\!E)_{1a} \!\uparrow\! G \oplus (B)_{1a} \!\uparrow\! G \oplus (A)_{2c} \!\uparrow\! G$.
	
	Similarly, the square-square-triangle-triangle-triangle-triangle lattice has the $1a$ and $1b$ positions occupied by faces, and the $2c$ position occupied by an edge.
	The line-graph lattice RSIs are then $\delta^4_{1a, 1} = \delta^4_{1a, 2} = \delta^4_{1b, 1} = \delta^4_{1b, 2} = 0$ and $\delta^2_{2c, 1} = -1$.
	Here there are multiple equivalent solutions for the flat-band representation, one of which is $2(A)_{2c} \!\uparrow\! G \oplus (B)_{2c} \!\uparrow\! G$.
	
	We leave it to future work to prove these statements, generalize to all such line-graph lattices with $C_4$ symmetry, and examine how bands with fragile topology may be created through perturbation.

\end{document}